\documentclass[lettersize,journal,compsoc]{IEEEtran}
\usepackage{amsmath,amsfonts}
\usepackage{array}
\usepackage[caption=false,font=normalsize,labelfont=sf,textfont=sf]{subfig}
\usepackage{textcomp}
\usepackage{stfloats}
\usepackage{url}
\usepackage{verbatim}
\usepackage{graphicx}
\usepackage{cite}
\usepackage{times}
\usepackage{epsfig}
\usepackage{graphicx}
\usepackage{amsmath}
\usepackage{amssymb}

\usepackage{booktabs}
\usepackage{float}  
\usepackage{multirow}
\usepackage{caption}
\usepackage{soul}
\usepackage{dsfont}
\usepackage{enumitem}
\usepackage{marvosym}
\usepackage{color}
\usepackage{amsthm}
\usepackage{pifont}
\usepackage{cleveref}
\usepackage[table]{xcolor}
\begin{document}
\title{RepCaM++: Exploring Transparent Visual Prompt with Inference-time Re-parameterization for Neural Video Delivery}

\author{Rongyu Zhang,~\IEEEmembership{Student Member,~IEEE}, Xize Duan, Jiaming Liu,~\IEEEmembership{Student Member,~IEEE}, \\ Li Du,~\IEEEmembership{Member,~IEEE}, Yuan Du,~\IEEEmembership{Senior Member,~IEEE}, Dan Wang \Letter,~\IEEEmembership{Senior Member,~IEEE}, \\ Shanghang Zhang \Letter, ~\IEEEmembership{Member,~IEEE}, Fangxin Wang,~\IEEEmembership{Member,~IEEE}

\thanks{This paper is an extended version of \cite{zhang2023repcam}, in Proceedings of the 33rd Workshop on Network and Operating System Support for Digital Audio and Video.}
\thanks{Corresponding author \Letter: Dan Wang and Shanghang Zhang.}
\thanks{Rongyu Zhang is with the National Key Laboratory for Multimedia Information Processing, School of Computer Science, Peking University. He is also a dual Ph.D. student with both Nanjing University and The Hong Kong Polytechnic University. (e-mail: rongyuzhang@smail.nju.edu.cn).}
\thanks{Xize Duan and Fangxin Wang are with the Future Network of Intelligence Institute (FNii) and the School of Science and Engineering (SSE), The Chinese University of Hong Kong, Shenzhen (e-mail: xizeduan@link.cuhk.edu.cn; wangfangxin@cuhk.edu.cn).}
\thanks{Jiaming Liu and Shanghang Zhang are with the National Key Laboratory for Multimedia Information Processing, School of Computer Science, Peking University (email: jiamingliu@stu.pku.edu.cn; shanghang@pku.edu.cn).}
\thanks{Yuan Du and Li Du are with Nanjing University (email: yuandu@nju.edu.cn; ldu@nju.edu.cn).}
\thanks{Dan Wang is with the Department of Computing, The Hong Kong Polytechnic University, Hong Kong (e-mail: csdwang@comp.polyu.edu.hk).}
}

\markboth{IEEE TRANSACTIONS ON MOBILE COMPUTING, VOL., NO., 2024}%
{Shell \MakeLowercase{\textit{et al.}}: A Sample Article Using IEEEtran.cls for IEEE Journals}

\IEEEtitleabstractindextext{
\begin{abstract} 
Recently, content-aware methods have been employed to reduce bandwidth and enhance the quality of Internet video delivery. These methods involve training distinct content-aware super-resolution (SR) models for each video chunk on the server, subsequently streaming the low-resolution (LR) video chunks with the SR models to the client. Prior research has incorporated additional partial parameters to customize the models for individual video chunks. However, this leads to parameter accumulation and can fail to adapt appropriately as video lengths increase, resulting in increased delivery costs and reduced performance. In this paper, we introduce RepCaM++, an innovative framework based on a novel Re-parameterization Content-aware Modulation (RepCaM) module that uniformly modulates video chunks. The RepCaM framework integrates extra parallel-cascade parameters during training to accommodate multiple chunks, subsequently eliminating these additional parameters through re-parameterization during inference. Furthermore, to enhance RepCaM's performance, we propose the Transparent Visual Prompt (TVP), which includes a minimal set of zero-initialized image-level parameters (e.g., less than 0.1\%) to capture fine details within video chunks. We conduct extensive experiments on the VSD4K dataset, encompassing six different video scenes, and achieve state-of-the-art results in video restoration quality and delivery bandwidth compression.
\end{abstract} 

}
\maketitle

\section{Introduction}
In contemporary digital ecosystems, mobile and edge computing have become crucial for accessing streaming media on smartphones and tablets, facilitating ubiquitous and uninterrupted access to high-resolution (HR) content. This shift has underscored the complexities of video streaming, where higher resolutions significantly strain already limited bandwidth capacities, especially within mobile network infrastructures. Advanced research \cite{zhang2023repcam,xiao2024task,wang2024bandwidth,chen2022macrotile} has focused on addressing these bandwidth constraints to improve the efficiency of video transmission.
These academic efforts chart a path toward a transformative transmission schema, wherein servers relay low-resolution (LR) video streams and dispatch super-resolution (SR) models to client devices. This strategy marks a significant departure from conventional methodologies employed in single image SR~\cite{shi2016real,dong2014learning,lim2017enhanced,zhang2018image,wang2022adaptive} and video SR~\cite{caballero2017real,chan2020basicvsr,chan2021basicvsr++}, by leveraging the unique overfitting property inherent in neural networks to enhance video delivery.

\begin{figure*}[t]
\includegraphics[width=\textwidth]{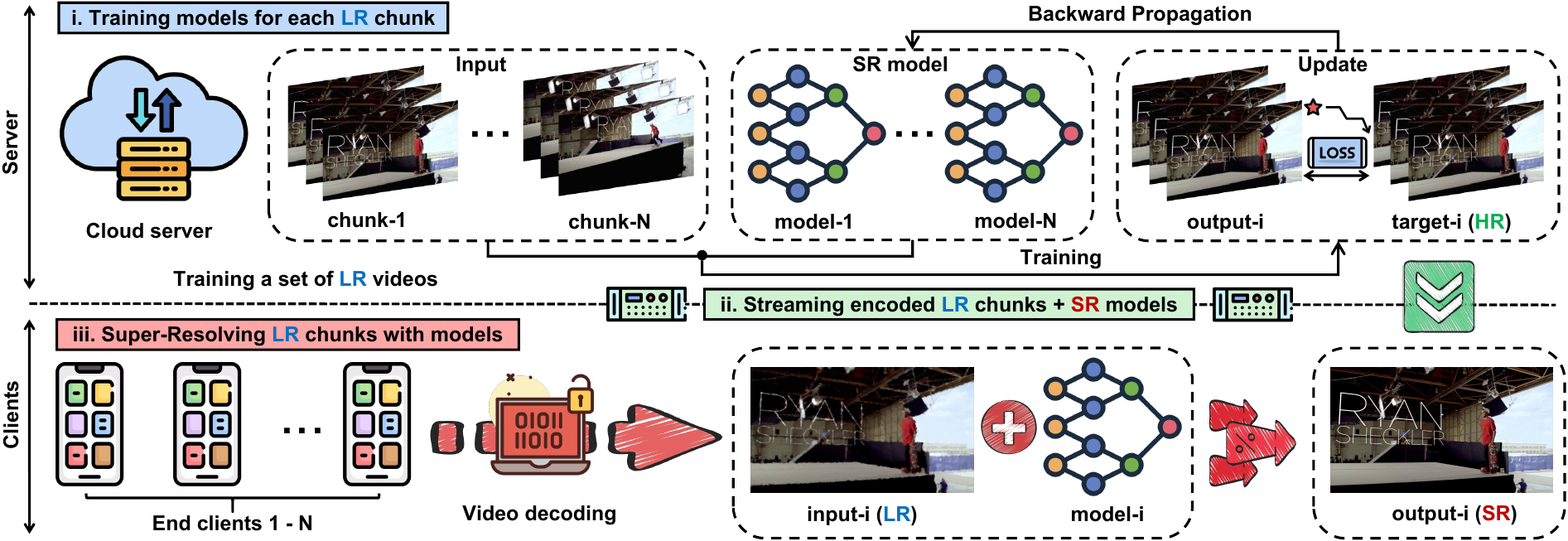}
\centering
\caption{\textbf{The whole procedure of adopting content-aware DNNs for video delivery.} A video is first divided into several chunks, and the server trains one model for each chunk. Then, the server delivers LR video chunks and models to the clients. The client runs the inference to super-resolve the LR chunks and obtain the SR video.}
\label{fig:teaser}
\end{figure*}

In this approach, videos are partitioned into $N$ discrete chunks, each associated with a distinctively overfitted SR model of size $S$ on the server side. Once transferred, client devices run these models to upgrade LR chunks of size $L$ into high-resolution ones. This process, termed Neural Video Delivery (NVD), ensures high-quality video content delivery under limited bandwidth conditions, particularly in mobile computing scenarios. The entire process is graphically depicted in \cref{fig:teaser}, offering a comprehensive illustration of the methodology for superior video transmission fidelity.


The promise of current research in SR models is undeniable. Yet, their application to NVD presents inherent challenges: \ding{202} Traditional methods frequently incur substantial bandwidth costs computed as $N\times (S+L)$ since multiple SR models corresponding to different video chunks must be transmitted from servers to clients. Consequently, the overhead of transferring SR models can sometimes exceed sending the HR videos directly. \ding{203} Advanced research~\cite{liu2021overfitting,li2022efficient} attempts to enhance the perception ability of the SR model with additional involved parameters to overfit the complete video. However, these approaches assume temporal homogeneity in video frames that weakens with longer content and varied scene dynamics.


\begin{figure*}[t]
\includegraphics[width=\textwidth]{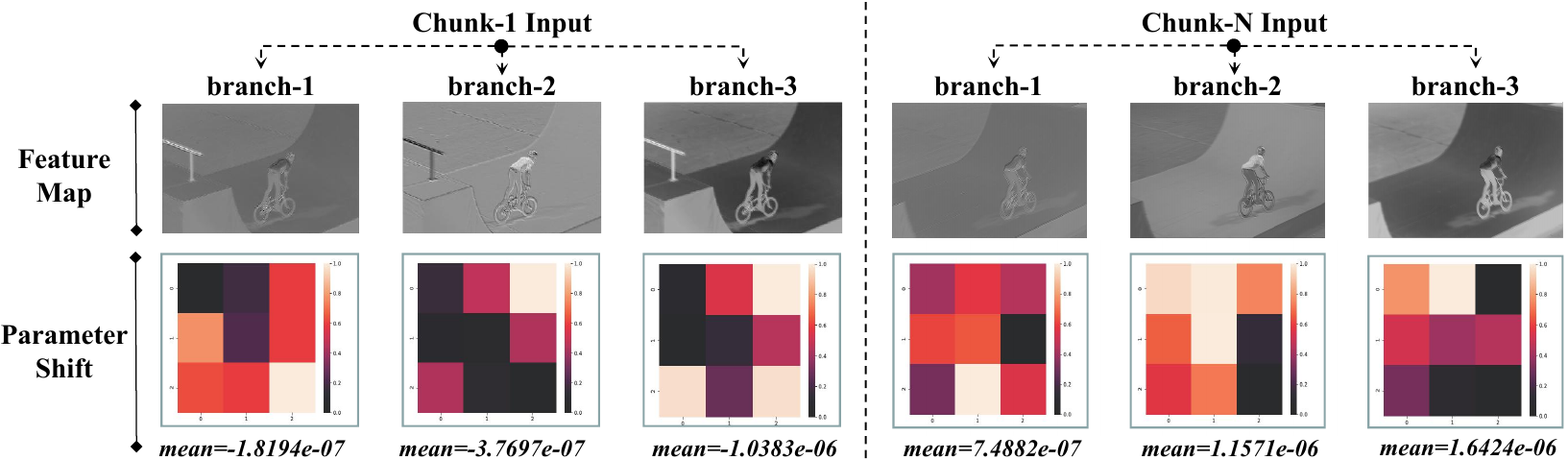}
\centering
\caption{Motivation of Re-parameterization Content-aware Modulation. (a) First row: we visualize the feature maps from different branches in the same position of the network (e.g. the third Conv layer) on randomly selected I frames from chunk-1 and chunk-n. (b) Second row: we calculate the gradient variation of different branches after the loss penalty and visualize the parameter sensitivity heat map in the second row. Different branches show different feature representations and perception abilities on the same frame.}
\label{fig:mot}
\end{figure*}

To overcome the aforementioned challenge \ding{202}, we propose a cutting-edge method, known as Re-parameterization Content-aware Modulation (RepCaM), to uniformly modulate entire video frames via an end-to-end training paradigm with a single model. Building on the observation by \cite{liu2021overfitting} that the relation of a certain feature map between SR models of different chunks is linear, RepCaM incorporates the linear-based, parallel-cascade convolutional layers into the SR model backbones as additional branches. This multi-branch architecture enables RepCaM to capture the linear relationships of the video features within a single SR model, thus eliminating the need for multiple models in NVD and enhancing efficiency during training, as illustrated in \cref{fig:mot}. During inference, we can consolidate the expanded branch network back to its original backbone form through re-parameterization~\cite{ding2021repvgg} without any model performance loss. Consequently, RepCaM facilitates overfitting an entire video with a single SR model, significantly reducing transmission costs to merely $S + N \times L$.


However, using the entire video as the sole input for RepCaM can still lead to a natural decrease in model overfitting as long video content changes over extended durations, as mentioned in challenge \ding{203}. Therefore, adding a module that can distinctly perceive specific differences in each video section emerges as a solution. As visual prompt (VP) has proven to be a cost-effective strategy for adapting large-scale Transformer models to various computer vision tasks, it adds less than 1\% of trainable parameters to the input space while maintaining the model's fundamental architecture towards enhanced model performance. However, traditional visual prompt~\cite{gan2023decorate,yang2023exploring} are generally initialized randomly to explore the parameter space more broadly and avoid local minima towards generalization, which is the opposite optimization direction compared to our overfitting model optimization goal. Moreover, directly adopting traditional VP into RepCaM will block parts of the input images and render them invisible. Thus, the blocked parts are treated as complete noise for our content-aware model and cannot well-overfit the original pixel values.

To overcome the aforementioned challenges and improve the ability of the RepCaM to handle long, varying video sequences, we introduce a novel and extremely light-weight Transparent Visual Prompt (TVP) of size $T$, which is zero-initialized with only hundreds of parameters and detailed in \cref{fig:prompt}. Our TVP embeds a latent instructive signal into the video frame, enriching the input image to better support SR models in identifying essential features and subtle differences within video sequences. Different from traditional random-initialized VP methods, our proposed TVP preserves the integrity of the original image inputs. It ensures that only the most significant changes are emphasized, allowing for focused attention on critical regions within long video sequences. Additionally, while zero initialization in TVP can cause nearly synchronized neuron updates and potentially limit generalization in most vision tasks, this feature proves beneficial in SR tasks by improving the model's resistance to overfitting. To further reduce bandwidth, we apply a single TVP across multiple frames to cut redundancy. Integrating this TVP with video chunks and sending it to clients with RepCaM, we establish the RepCaM++ network and greatly enhance video super-resolution at a transmission cost of $S+N\times(L+T)$.

In our study, we extensively evaluated RepCaM++ using the VSD4K video streaming dataset, including six different scenes, demonstrating its superiority over state-of-the-art baselines in terms of super-resolution quality PSNR and communication efficiency. Additionally, our method represents an innovative approach to video coding, extending its utility beyond mere content delivery. Compared with H.264 and H.265 codecs methods at the same storage cost, our approach attained a higher PSNR, attributed to its overfitting property, indicating its considerable promise.
Our contributions can be concluded as follows:
\begin{itemize}
\item We propose the innovative framework \texttt{RepCaM++} for NVD with an end-to-end training strategy and only minimal computational and storage overhead (under 0.1\%).
\item We introduce the RepCaM module to enhance the model perception and overfitting ability with a multi-branch architecture, which can be consolidated to maintain transmission and inference efficiency without performance loss.
\item We devise a novel TVP module to further substantially bolster the perceptual capabilities of the RepCaM for long videos, utilizing an exceedingly constrained parameter augmentation towards state-of-the-art performance.
\end{itemize}

\section{Related work}

{\bf DNN-based Super-Resolution.} 
SRCNN~\cite{dong2014learning} pioneered the use of DNNs for Super-Resolution, sparking a wave of research to enhance Single Image Super-Resolution (SISR). VDSR~\cite{kim2016accurate} shifted the focus to predicting high-frequency image residuals, employing a deep network architecture. SRResNet~\cite{ledig2017photo} further refined this approach by incorporating Residual Blocks~\cite{he2016deep}, which significantly improved SR performance. An insightful study by~\cite{lim2017enhanced} on the role of batch normalization in SR tasks led to its removal in SRResNet, yielding performance gains.
As attention mechanisms gained traction, RCAN~\cite{zhang2018image} became the first to integrate them into SR networks, allowing for deeper and more capable models, albeit with increased computational complexity. To balance efficiency and effectiveness, methods like~\cite{shi2016real} introduced pixel-shuffle layers to upscale low-resolution inputs directly, while~\cite{li2020lapar} pursued a linearly-assembled adaptive regression strategy.
It's important to note that these image SR techniques train on extensive databases such as DIV2K~\cite{Agustsson_2017_CVPR_Workshops} and apply to static images, overlooking the overfitting property inherent in DNNs that could be advantageous in video streaming.

\noindent{\bf Neural Video Delivery.} Neural Adaptive Streaming (NAS)~\cite{yeo2018neural} is an innovative Internet video delivery framework employing Deep Neural Networks for quality enhancement, adept at mitigating video quality degradation in bandwidth-constrained environments. NAS demonstrates the ability to either enhance the average Quality of Experience (QoE) by 43.0\% within the same bandwidth constraints or to economize bandwidth usage by 17.13\% while maintaining equivalent QoE levels. The cornerstone of NAS is the strategic utilization of DNNs' overfitting property, harnessing training accuracy to amplify Super-Resolution performance. This paradigm has been extended in various domains, including UAV video streaming~\cite{xiao2019sensor}, live streaming~\cite{kim2020neural}, generic video streaming~\cite{dasari2020streaming, chen2020sr360}, volumetric video streaming~\cite{zhang2020mobile}, and mobile video streaming~\cite{yeo2020nemo}. Recent advancements~\cite{liu2021overfitting,khani2021efficient} have sought to further minimize bandwidth consumption by sharing parameters across video chunks, transmitting only a minimal set of unique parameters per chunk. The STDO~\cite{li2023towards} method capitalizes on spatial-temporal characteristics to chunk videos intelligently, thereby minimizing the number of chunks and associated model dimensions. EMT~\cite{li2022efficient} leverages meta-learning to adapt a global model to the initial video chunk, followed by selective fine-tuning of parameters informed by gradient activity in previously adapted models. Despite significant advancements, existing methods fall short of addressing the computational demands of network training and the nuanced dynamics of scene transitions for optimal video chunking. These challenges remain pivotal for improving the feasibility and efficiency of video streaming technologies. Our work addresses these issues by integrating techniques into a unified model that overfits through data-aware joint training, significantly cutting storage requirements with negligible quality degradation.

\noindent{\bf Visual Prompt Tuning.} 
Visual Prompt Tuning (VPT)~\cite{jia2022visual,bar2022visual,bahng2022exploring} has arisen as a resource-conscious alternative to exhaustive fine-tuning of large-scale Transformer models for computer vision tasks, integrating a marginal quantity of tunable parameters (less than 1\% of the total) into the input space while preserving the core architecture. Recent studies have extensively applied visual prompts to domain adaptation and out-of-distribution scenarios. DePT~\cite{gao2022visual} incorporates visual prompts into vision Transformers, fine-tuning only these initialized prompts during the adaptation process. Gan et.al.~\cite{gan2023decorate} developed domain-specific and domain-agnostic prompts for extracting pertinent domain knowledge while retaining commonalities across domains during continuous adaptation. Yang et al.~\cite{yang2023exploring} introduced Sparse Visual Domain Prompts (SVDP), a technique that conserves input spatial information and augments domain-specific knowledge extraction by selectively allocating tunable parameters to pixels experiencing significant distribution shifts. In image processing, Bar et.al.~\cite{bar2022visual} demonstrated task adaptation using pre-trained visual models without task-specific fine-tuning or architectural alterations by using a visual prompt image with an integrated placeholder. Liu et al.~\cite{liu2023explicit} introduced Explicit Visual Prompting (EVP), a model that directs tunable parameters to explicitly focus on the visual content of each image, namely the features from static patch embeddings and high-frequency components of the input. Nonetheless, the application of visual prompts to SR models remains unexplored.

\noindent{\bf Structural Re-parameterization.} Recent work has highlighted the benefits of reparameterizing skip connections to reduce memory access costs. For instance, RepVGG~\cite{ding2021repvgg} introduced a simple yet effective convolutional neural network architecture. During inference, this architecture comprises solely a stack of 3$\times$3 convolutional layers, whereas the training-time model features a multi-branch topology. Building on RepVGG, MobileOne~\cite{vasu2023mobileone} implemented trivial over-parameterization in its branches, achieving further accuracy improvements. Chu et al.~\cite{chu2024make} advanced this concept by developing a simple, robust, and effective solution to create a quantization-friendly structure that also benefits from re-parameterization. Following recent advancements in Transformer~\cite{vaswani2017attention} and Vision Transformer~\cite{dosovitskiy2020image}, RepVit~\cite{wang2024repvit} augmented the mobile-friendliness of standard lightweight CNNs by incorporating the efficient architectural designs of lightweight Vision Transformers. Additionally, FastVit~\cite{vasu2023fastvit} applied train-time over-parametrization and large kernel convolutions, demonstrating that these choices significantly enhance accuracy while having minimal impact on latency. While previous works have primarily focused on improving model efficiency for general vision tasks, RepCaM~\cite{zhang2023repcam} and our work is pioneering the application of the re-parameterization technique in the video delivery domain. This novel application aims to reduce bandwidth consumption and improve video quality, marking a significant extension of re-parameterization benefits beyond traditional areas.

\section{Proposed methods}
\label{PM}
In this section, we detail \texttt{RepCaM++} to enhance NVD. The \cref{motivation} covers observations on parallel-cascade branch architectures and TVP. The \cref{training} illustrates the RepCaM, and \cref{convert} explains inference consolidated through Re-parameterization~\cite{ding2021repvgg}. Finally, the \cref{TVP} details the design of TVP, which, combined with RepCaM to form the RepCaM++.

\begin{figure*}[t]
\includegraphics[width=0.97\textwidth]{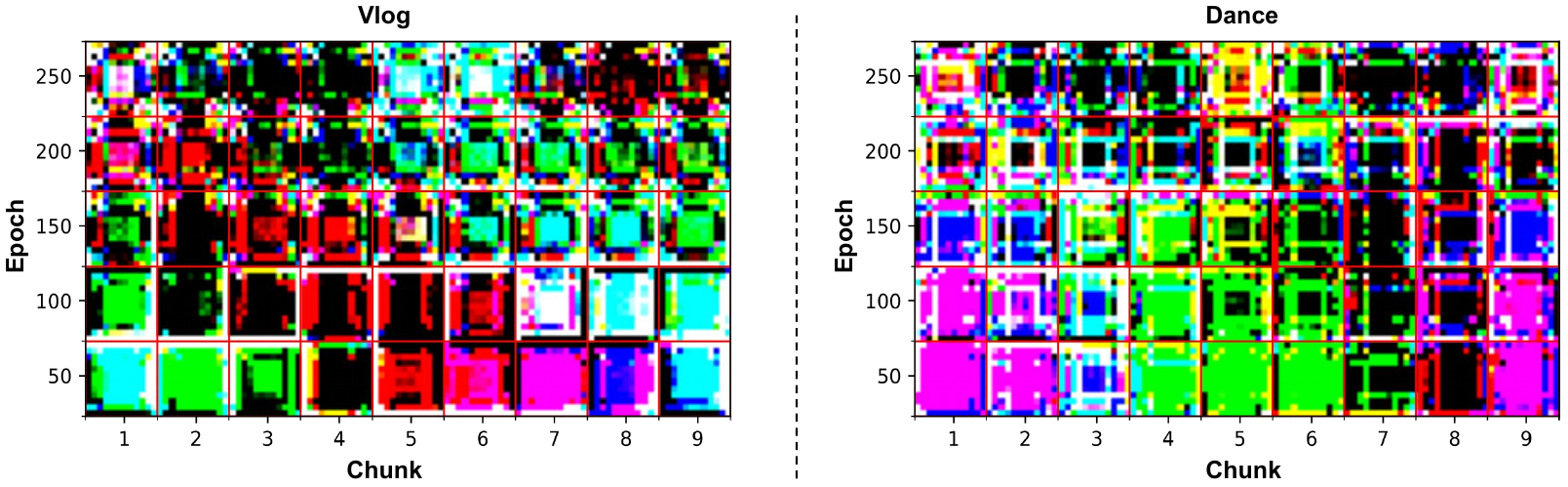}
\centering
\caption{Visualization of the training progression for Transparent Visual Prompt (TVP): (1) Vertically, it adapts to specific features. (2) Horizontally, different TVPs preserve the temporal information across various chunks. For enhanced clarity, we scale the pixel values from the range (0,1) to (0,255).}
\label{fig:prompt}
\vspace{-0.3cm}
\end{figure*}

\subsection{Motivations}
\label{motivation}
\textbf{Re-parameterization Content-aware Modulation.} In this section, we articulate the motivation for the Re-parameterization Content-aware Modulation (RepCaM) module.
Specifically, in alignment with preceding studies~\cite{kim2020neural, yeo2018neural}, we chunk a 45-second video from the VSD4K dataset \cite{liu2021overfitting} into $n$ portions. Then, we utilize a network with multiple parallel-cascade 3$\times$3 convolutional branches, each with varying numbers of 1$\times$1 convolutional layers. We conduct simultaneous training on separate chunks to explore the distinct feature representation capabilities of each branch.

As depicted in the first row of \cref{fig:mot}, the visualization of feature maps corresponding to each branch yields an empirical finding: different branches in a linear relationship show unique feature representation to the same input. Subsequently, we engage in the quantification of gradient fluctuations across these branches. The resulting parameter sensitivity is graphically rendered as heat maps in the second row of \cref{fig:mot}, from which it is apparent that each branch possesses a unique degree of sensitivity to variant inputs. This diversity in feature and gradient sensitivity helps the model to better understand and process video content by capturing a wider range of features.

In addition, Liu et.al.~\cite{liu2021overfitting} have revealed that the relation of a certain feature map among SR models corresponding to each different video chunk is linear. The insights gleaned from the observations and conclusion have prompted us to adopt a parallel-cascade network with an inherently established linear relationship to replace multiple SR models for transmission cost saving. Moreover, such architecture further allows for different levels of implicit modulation of the same individual video frame. Furthermore, the added parameters that increase the model's delivery overhead motivate us to apply re-parameterization techniques~\cite{ding2021repvgg} and propose the RepCaM. RepCaM takes advantage of a multi-branch structure during the training and consolidates the network during inference, thus improving video quality without incurring extra communication costs and tackling a key bottleneck in NVD.

\noindent\textbf{Transparent Visual Prompt.} Visual prompts have been widely applied to various vision tasks to enhance model generalization, such as in domain adaptation~\cite{gan2023decorate,yang2023exploring}, which contradicts the optimization goal for neural video delivery. Moreover, traditional VP is generally randomly initialized and obscure chunks of the input imagery, rendering these sections imperceptible and resistant to overfitting by neural networks. To overcome the aforementioned challenges and leverage visual prompts effectively, we propose the TVP with zero-initialization, which is naturally prone to overfitting and makes them transparent for input videos.

However, we are also curious to determine whether the TVP can efficiently perceive the various input features and not fail in the symmetry parameter update. Therefore, similar to our previous exploration, we divide the videos into 9 chunks, place 9 distinct 48 $\times$ 48 TVPs at the center of each video chunk, and train on Vlog and Dance data from the VSD4K dataset. We visualize these changes throughout the training on every chunk. As illustrated in \cref{fig:prompt}, the TVPs update alongside the propagation of image tensors and adapt to various scenarios. Vertically, we observe that TVPs gradually learn from the ever-changing distribution of input samples across different epochs within each chunk. Horizontally, TVPs show different levels of consistency across different chunks, indicating their ability to preserve temporal information across various chunks. Therefore, we can conclude that the collection of TVPs can learn the unique distributions in different video sequences while maintaining a commonality toward data overfitting.

\vspace{-0.3cm}
\subsection{Re-parameterization content-aware modulation}
\label{training}
\begin{figure*}[t]
\includegraphics[width=0.95\textwidth]{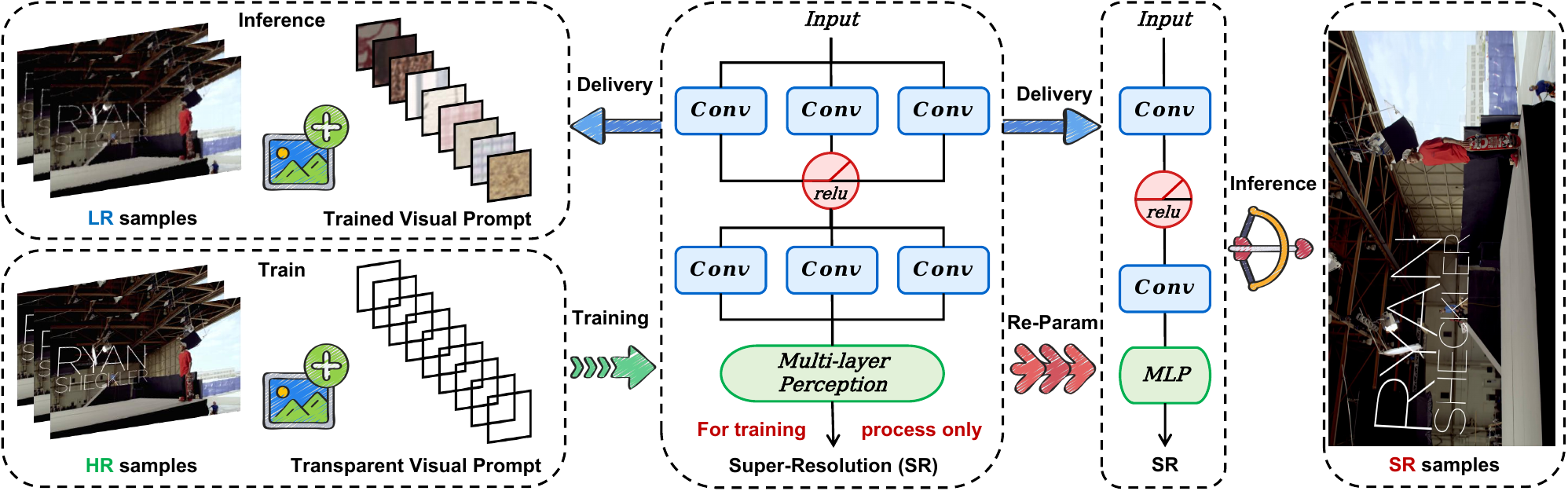}
\centering
\caption{The overall framework of our proposed RepCaM++, consisting of RepCaM and TVP, only requires delivering the LR chunks with one shared re-parameterized model, while the original super-resolution model will only be used for training.}
\label{fig:comparison}
\end{figure*}

\begin{figure}[t]
\includegraphics[width=0.48\textwidth]{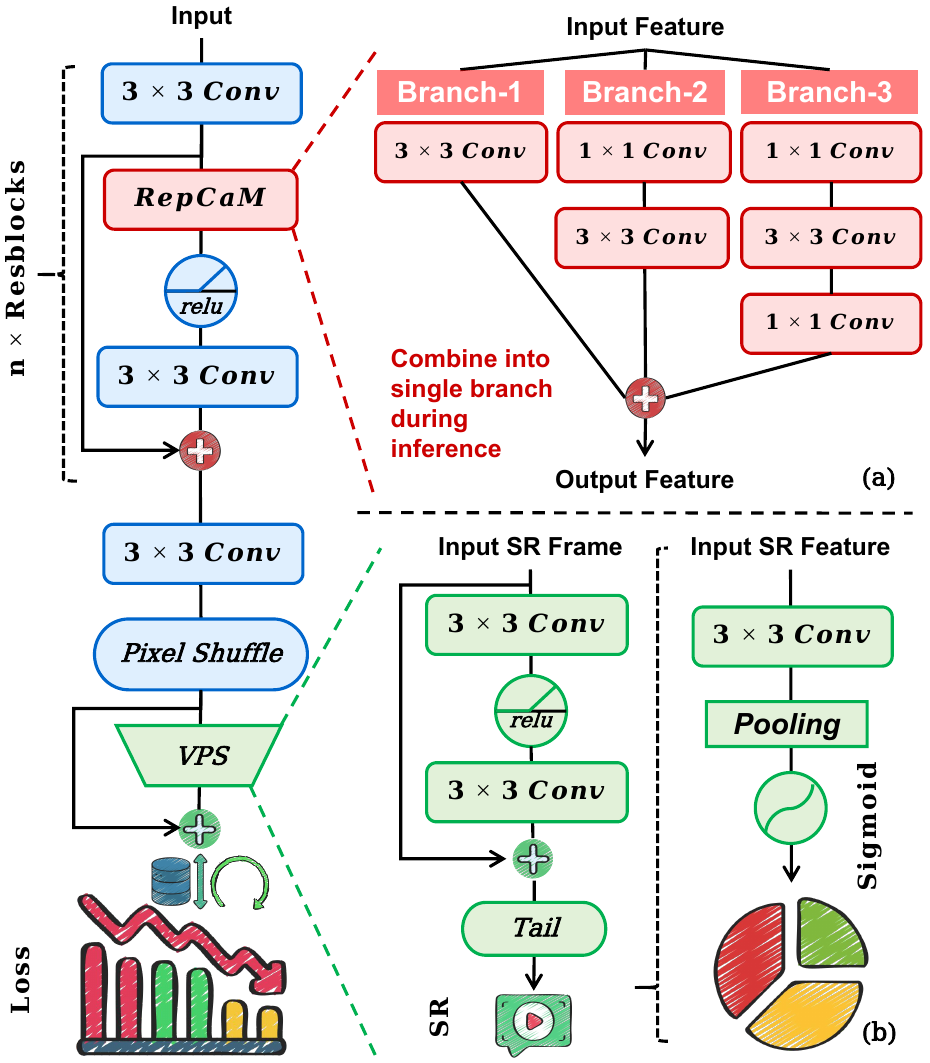}
\centering
\caption{The network architecture of our proposed RepCaM module. Part (a) is the parallel-cascade RepCaM network. Part (b) aims to realize the VSP function.}
\label{fig:net}
\end{figure}

In this section, we present the design of the RepCaM network. This approach stems from insights discussed in \cref{motivation}, leading us to integrate parallel-cascade learnable parameters within the content-aware training paradigm. As shown in \cref{fig:net}(a), our multi-branch architecture functions as an ensemble of multiple simplified models~\cite{veit2016residual}, specifically an ensemble of $2^{n}$ permutations from $n$ residual blocks. This design expands the parameter space of the basic residual module~\cite{lim2017enhanced}, enhancing both breadth and depth. This expansion significantly boosts the modulation capacity and perceptual quality of the RepCaM model. To highlight our advancements, we detail our framework structure in \cref{fig:comparison}.

To elucidate our approach, we have restructured the traditional Conv layer within the residual module, adopting a horizontally oriented tri-branch configuration. Each branch is outfitted with a $3\times3$ Conv layer, selected specifically for its comparative advantages in modulation capacity and computational efficiency, exemplified by its optimal theoretical TFLOPS of 38.10 \cite{ding2021repvgg}. This choice is predicated on the fact that $3\times3$ Conv layers offer a better trade-off in performance metrics compared to their larger $5\times5$ and $7\times7$ counterparts.

Expanding upon this, we incorporate varying quantities of $1\times1$ Conv layers within each branch to facilitate nuanced content modulation. This allows for a more comprehensive and diverse feature representation as each branch becomes adept at capturing intricate details within the data. The synergy of these multiple branches, each composed of a strategic combination of $1\times1$ and $3\times3$ Conv layers, constitutes what we term the RepCaM convolution. This configuration is not only more content-aware but also provides a richer representational capacity. We let input video frame represents $I\in \mathbb{R}^{H\times W}$ and the feature represents $x\in \mathbb{R}^{B\times C\times H\times W}$. The mathematical formulation of the RepCaM convolution is as follows:
\begin{equation}
 y=\sum^{M-1}_{i=0}(f(g_{i}(x))), \quad2\leq M
\end{equation}
wherein $M$ denotes the number of parallel branches, $f(\cdot)$ represents the convolutional operation with a $3\times3$ kernel, and $g_{i}(\cdot)$ signifies a sequence of $i$ consecutive $1\times1$ convolutional operations. Consequently, the RepCaM framework may be conceptualized as an amalgamation of 'multiple models', each branch functioning as a distinct model within the ensemble. This architectural strategy affords a substantial enhancement in the model's capacity to approximate complex functions, thereby offering the remarkable potential for accurately representing a greater diversity of video frame data \cite{yeo2018neural,liu2021overfitting}.

However, incorporating parallel-cascade learnable parameters has consequentially escalated the computational demands of training. Moreover, the inherent characteristics of video content induce temporal redundancy across successive frames. To address these challenges, drawing inspiration from existing literature \cite{sun2020learning,wang2021samplingaug}, we introduce a novel, efficient training methodology tailored for extended video sequences termed Video Patch Sampling (VPS).

As depicted in \cref{fig:net}(b), VPS employs a rudimentary classification sub-network to generate a weight map during the training phase. This map is then employed to adaptively select the most salient patches through a dot product operation with the SR video frames. The VPS strategy empowers the model to focus selectively on the most informative samples throughout the RepCaM training process. This targeted learning approach not only bolsters the model's precision but also significantly accelerates network convergence on computational servers.

\subsection{RepCaM during inference-time}
In this section, we detail the process for compressing the RepCaM model to match the parameter footprint of the original SR model. The compression pipeline is structured into two primary operations: cascade conversion, which sequentially realigns the multi-branch framework of RepCaM, and parallel concatenation, which consolidates the parameters into a singular, cohesive model structure. This approach ensures that the refined RepCaM model retains the performance benefits while conforming to the size constraints of the baseline SR architecture.

\noindent\textbf{Cascade sequential conversion.}
\label{convert}
In pursuit of architectural simplification, our objective is to consolidate the $1\times1-3\times3$ conv pair present within each branch into a singular $3\times3$ Conv layer. Drawing from the established methods in SR models \cite{shi2016real, lim2017enhanced}, where the input and output channels are typically equivalent, absent the extreme (first and last) Conv layers, we adopt $C$ to denote the channel dimension for both input and output of a Conv layer. We define the fourth-order tensor $Q^{(K)} \in \mathbb{R}^{C_{1} \times C_{2} \times K \times K}$, augmented by a bias term $b \in \mathbb{R}^{C_{1}}$, to encapsulate the parameters of a $K\times K$ Conv layer. The output of a $1\times1-3\times3$ Conv block, employing the kernel $Q^{(K)}$ configured in the dimensions of $C\times C\times K\times K$, with $K$ representing the kernel size, is formulated as follows:
\begin{equation}
F_{out}=(F_{in}\circledast Q^{(1)}+B(b^{(1)}))\circledast Q^{(3)}+B(b^{(3)})
\end{equation}
where $\circledast$ represents the convolution operation. The variables $F_{in}$ and $F_{out}$, both members of the set $\mathbb{R}^{C\times H\times W}$, correspond to the input and output feature maps, respectively. The function $B(\cdot)$ expands the bias vector $b$ into a bias matrix $B(b^{(K)}) \in \mathbb{R}^{C\times H'\times W'}$, where the dimensions $H'$ and $W'$ are contingent upon the specific convolutional parameters. Invoking the principle of convolution linearity \cite{ding2021diverse}, we can derive the requisite formulation for the kernel and bias of an equivalent single convolution layer, denoted $Q'$ and $b'$, respectively:
\begin{equation}
F_{out}=F_{in}\circledast Q'+B(b')
\end{equation}
Consequently, a $3\times3$ convolutional layer superimposed with an arbitrary sequence of $1\times1$ convolutions can be effectively condensed into a solitary $3\times3$ convolutional layer. It is imperative to emphasize that for this compression to be feasible, the convolutional layers in question must share identical configurations, including padding, stride, and other pertinent parameters.

\noindent\textbf{Parallel branches concatenation.}
\label{concat}
Upon transforming the cascade of convolutional blocks into a single $3\times3$ convolutional layer, we proceed to the branch combination phase utilizing concatenation. Two branches, characterized by their respective convolutional kernels $Q^{(3)}_{1} \in \mathbb{R}^{C\times C\times 3\times 3}$ with bias $b^{3}_{1} \in \mathbb{R}^{C}$, and $Q^{(3)}_{2} \in \mathbb{R}^{C\times C\times 3\times 3}$ with bias $b^{3}_{2} \in \mathbb{R}^{C}$, can be seamlessly merged. This is achieved by concatenating the kernels along the channel dimension to form $Q' \in \mathbb{R}^{(C+C) \times C \times 3 \times 3}$, accompanied by a corresponding concatenated bias vector $b' \in \mathbb{R}^{C+C}$:
\begin{equation}
\begin{aligned}
CONCAT(F_{in}\circledast Q^{(3)}_{1}+B(b^{(3)}_{1}), &F_{in}\circledast Q^{(3)}_{2}+B(b^{(3)}_{2})) \\
=&F_{in}\circledast Q'+B(b')
\end{aligned}
\end{equation}

Given that concatenation effectively amalgamates convolutional layers along their output channel axis, and considering our $3\times3$ convolutions across each branch are configured identically, the multi-branch structure can coalesce into a unified $3\times3$ convolutional layer. This consolidation leverages the homogeneity in layer configurations to facilitate seamless integration, resulting in a streamlined architecture.




\subsection{Transparent visual prompt}
\label{TVP}
The processing of entire video frames as a monolithic input can be problematic due to the volatile attributes of foreground components. To address this, we partition the video into individual chunks and employ a novel construct known as Transparent Visual Prompt, which significantly bolsters the model's interpretative prowess of diverse scenes, with a negligible increase in transmission overhead—less than 0.1\%. In our super-resolution pipeline, each frame $I^{T}\in \mathbb{R}^{H\times W}$ within a chunk is pre-processed by superimposing a zero-initialized TVP, denoted as $w_{\phi}\in \mathbb{R}^{S_{H}\times S_{W}}$, which is spatially added in an element-wise fashion. The TVP is designed to be initially transparent, having a null effect on the native frame content, thereby ensuring fidelity in the initial super-resolution stages. During subsequent reconstruction, $w_{\phi}$ imparts pivotal supplementary data, adaptively amplifying the model's super-resolution capabilities. 
The TVP is formally defined as a learnable parameter matrix $w_\phi \in \mathbb{R}^{S_H \times S_W \times C}$, where $\{S_H, S_W\}$ represent its dynamically adjustable spatial dimensions and $C$ corresponds to the channel depth of the input feature map $I^T \in \mathbb{R}^{H \times W \times C}$. The TVP operates through the following piecewise function:
\begin{equation}
I^{T}_{p}[i,j] = 
\begin{cases} 
I^{T}[i,j] + w_{\phi}\left[i - \Delta_H, j - \Delta_W\right], & \text{if } (i,j) \in \mathcal{R}_{\text{TVP}} \\
I^{T}[i,j], & \text{otherwise}
\end{cases}
\end{equation}
where the activation region $\mathcal{R}_{\text{TVP}}$ is determined by:
\begin{equation}
\mathcal{R}_{\text{TVP}} = \left[ \Delta_H, \Delta_H + S_H - 1 \right] \times \left[ \Delta_W, \Delta_W + S_W - 1 \right]
\end{equation}
where center alignment offsets $\Delta_H$ and $\Delta_W$ are computed as:
\begin{equation}
\Delta_H = \left\lfloor \frac{H - S_H}{2} \right\rfloor, \quad \Delta_W = \left\lfloor \frac{W - S_W}{2} \right\rfloor
\end{equation}

The dynamic scaling parameters $\{S_H, S_W\}$ are optimized through backpropagation, enabling automatic adaptation to diverse low-resolution video characteristics.
From a mathematical perspective, this procedure can be interpreted as the integration of a modifiable bias into the initial image representation, while the final loss function for RepCaM++ relative to the TVP is crafted to ensure high-fidelity image reconstruction and systematically derived as:
\begin{equation}
\begin{aligned}
\label{loss}
\mathcal{L}(I^{T}_{p}) = \frac{1}{N}\sum_{i=1}^{N}(I^{T}_{p}-I^{T}_{HR})
\end{aligned}
\end{equation}
where $I^{T}_{HR}$ represents the original high-resolution video frame. During back-propagation, we calculate the gradient of the loss $\mathcal{L}$ with respect to the TVP $w_{\phi}$ via the chain rule:
\begin{equation}
\begin{aligned}
\label{gradient}
\partial \mathcal{L}/\partial w_{\phi}[i,j] = \partial \mathcal{L}/\partial I^{T}_{p}[i,j],
\end{aligned}
\end{equation}
Here, the term $\partial \mathcal{L}/\partial I^{T}{p}[i,j]$ denotes the gradient of the loss function $\mathcal{L}$ with respect to the transformed input image $I^{T}_{p}$ at the spatial location $(i, j)$. Given that the transformation weight matrix $w{\phi}$ is applied element-wise to the input tensor $I^{T}$ via an additive operation, the partial derivative of the transformed input $I^{T}{p}$ with respect to $w{\phi}$ simplifies to a constant value of 1. This implies that for every element within the designated visual prompt region, which encompasses a $S_{H} \times S_{W}$ pixel area, the gradient $\partial I^{T}{p}/\partial w{\phi}[i, j]$ retains a value of 1. This constant gradient reflects the direct influence of $w{\phi}$ on $I^{T}_{p}$ without any modulation, which is indicative of an identity transformation within the specified region. The TVP weight $w_{\phi}$ can then be updated using gradient descent:
\begin{equation}
w'_{\phi}[i,j] = w_{\phi}[i,j] - \eta\times \frac{\partial \mathcal{L}}{\partial w_{\phi}[i,j]},
\end{equation}
where $w'_{\phi}[i,j]$ represents the updated TVP parameters, and $\eta$ is the learning rate used by the optimizer. As a result, even though the initial values of $w_{\phi}[i,j]$ are zero, during training, the gradients $\partial \mathcal{L}/\partial w_{\phi}[i,j]$ will adjust the values of $w_{\phi}[i,j]$, allowing them to contribute to the model's learning process.

Integrating transparent visual prompts with a content-aware modeling framework offers a robust strategy for local clients aiming to attain superior video super-resolution outcomes. This methodology markedly improves the super-resolution process's efficiency and efficacy, particularly when contending with video sequences characterized by dynamic content changes. It guarantees the preservation and refinement of image detail, thereby delivering an enriched high-definition visual experience that faithfully maintains the original authenticity of the video content.

\section{Experiments}
\label{experiment}
In this section, we conduct extensive experiments to substantiate the merits of our proposed methodologies. We apply our developed RepCaM++ framework to videos of varied scenes and duration, demonstrating the versatility of our approach. Detailed descriptions of the video datasets and our implementation strategies are provided in \cref{sec:setting}. The \cref{sec:main} presents a comparative analysis with other state-of-the-art baselines across diverse videos and SR models, revealing that RepCaM++ delivers superior video super-resolution quality while also minimizing transmission costs. In \cref{sec:abla}, we illustrate the results achieved by our unified SR model utilizing RepCaM++, which effectively capitalizes on the heterogeneous information density across various video chunks, thereby enhancing training performance. The \cref{sec:abla} details our ablation study on the crucial parameters employed in the RepCaM++ framework, examining aspects such as the number of video chunks, the scheduling of training data, and the processing of lengthy videos with multiple scene transitions.

\subsection{Experimental details}
\label{sec:setting}
\textbf{Dataset.} Previous super-resolution research primarily used datasets like Vimeo-90K~\cite{xue2019video} and REDS\cite{nah2019ntire}  only consist of neighboring frame sequences, which are not suitable for video delivery. To mitigate this, we employ VSD4K~\cite{liu2021overfitting}, featuring six categories: vlog, game, interview, city, sports, and dance, with a range of video durations including 15s, 30s, 45s, 1min, 2min, and 5min. HR videos are established at 1080p resolution, while LR videos are bicubically interpolated to suit various scaling factors.

\noindent\textbf{Baselines.} Since our method can be applied to different SR architectures, we conduct extensive experiments with the popular networks, including EDSR~\cite{lim2017enhanced}, ESPCN~\cite{shi2016real}, VDSR~\cite{kim2016accurate}, SRCNN~\cite{dong2015image} to demonstrate the generalization ability of our method. We further compare our method with the previous neural video delivery methods:
\begin{itemize}
\item awDNN~\cite{yeo2017will} tends to overfit the entire video using a single SR model.
\item NAS~\cite{yeo2018neural} pre-divides a video into multiple chunks and overfits each time-divided chunk with a separate SR model.
\item CaFM~\cite{liu2021overfitting} employs a single SR model equipped with a handcrafted module to overfit time-divided video chunks.
\item EMT~\cite{li2022efficient} utilizes a meta-learning model adapted to the first chunk of the video. For subsequent chunks, it fine-tunes specific parameters chosen through gradient masking from the previously adapted model.
\end{itemize}

\noindent\textbf{Training Details.} 
Following previews of neural video delivery works~\cite{liu2021overfitting, li2022efficient}, We train and test on the same video frames in a content-aware manner with scaling factors $\times2$, $\times3$, and $\times$4. We take $48\times 48$ HR patches with corresponding LR patches for training. The visual prompt is set to be $48\times 48$ and is added to the center of the input video features, and the videos are divided into 9 chunks. In order to reduce the computational cost, we sample 1 frame out of 10 frames for testing. Adam is selected as the optimizer with $\alpha=0.9$, $\alpha=0.999$, $\epsilon=10^{-8}$. We conduct our experiment on the EDSR model with 16 resblocks. We utilize L1 loss, and the learning rate is set as $5e-5$ and decays at 200 epochs. The evaluation metrics of PSNR, LPIPS, and SSIM provide a comprehensive assessment of image quality. PSNR quantifies reconstruction accuracy, LPIPS aligns with human visual perception, and SSIM evaluates structural preservation, collectively offering insights into the performance of neural video delivery. All our experiments are conducted on GeForce RTX 3090 GPU.

\begin{figure}[t]
\includegraphics[width=0.5\textwidth]{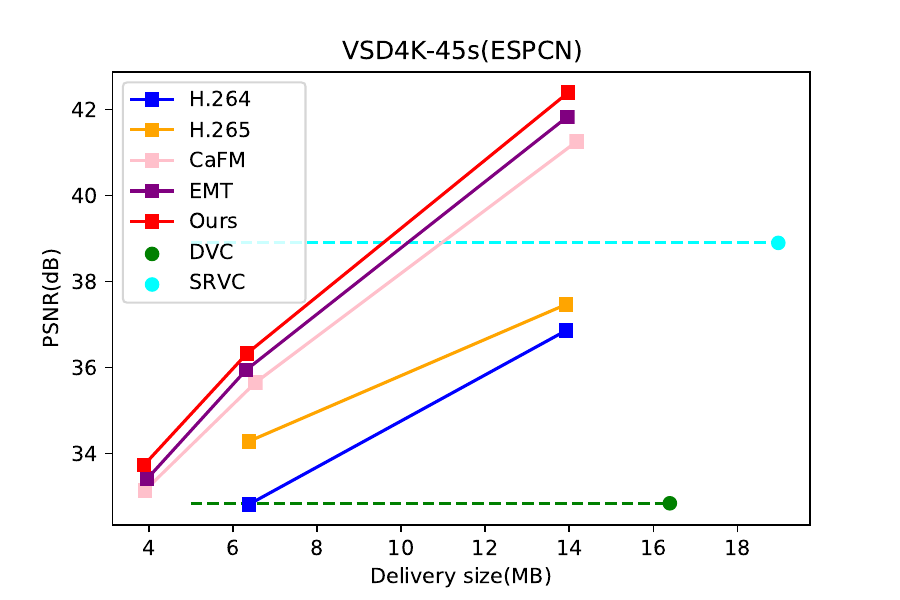}
\centering
\caption{Comparisons with neural video delivery methods in terms of PSNR and delivery size.}
\label{fig:size}
\vspace{-0.4cm}
\end{figure}

\begin{table*}[t]
\linespread{0.9}
\small
\centering
\caption{PSNR results of RepCaM++ on game, inter, and vlog VSD4K data with different SR backbones.}
\scalebox{0.925}{
\begin{tabular}{l|l|p{1.3cm}p{1.3cm}p{1.3cm}|p{1.3cm}p{1.3cm}p{1.3cm}|p{1.3cm}p{1.3cm}p{1.3cm}}
\toprule
&  & \multicolumn{3}{c}{\bf { game-15s-PSNR $\uparrow$ }}& \multicolumn{3}{c}{\bf { inter-15s-PSNR $\uparrow$ }} & \multicolumn{3}{c}{\bf {vlog-15s-PSNR $\uparrow$}}  \\
\bf { Model } & \bf { Method } & $\times$2 & $\times$3 & $\times$4 & $\times$2 & $\times$3 & $\times$4 & $\times$2 & $\times$3 & $\times$4 \\
\midrule 
\multirow{5}{*}{ESPCN} & awDNN~\cite{yeo2017will} & 37.94 & 32.85 & 29.97 & 40.43 & 35.36 & 29.91 & 46.41 & 42.90 &39.65 \\
& NAS~\cite{yeo2018neural} & 37.58 & 32.71 & 30.59 & 40.62 & 35.42 & 30.43 & 46.53 & 43.01 & 39.98 \\
& CaFM\cite{liu2021overfitting} & 38.07 & 33.14 & 30.96 & 40.71 & 35.54 & 30.47 & 47.02 & 43.20 & 40.16 \\
& EMT~\cite{li2022efficient}& 38.14 & 33.23 & 30.99 & 40.68 & 35.63 & 30.51 & 47.01 & 43.29 & 40.20 \\
& \cellcolor{green!10}\bf{RepCaM++} & \cellcolor{green!10}\textbf{38.44} & \cellcolor{green!10}\textbf{33.51} & \cellcolor{green!10}\textbf{31.31} & \cellcolor{green!10}\textbf{40.76} & \cellcolor{green!10}\textbf{35.85} & \cellcolor{green!10}\textbf{30.79} & \cellcolor{green!10}\textbf{47.35} & \cellcolor{green!10}\textbf{43.44} & \cellcolor{green!10}\textbf{40.36} \\
\midrule
\multirow{5}{*}{SRCNN} & awDNN~\cite{yeo2017will} & 36.08 & 31.94 & 29.90 & 40.46 & 33.95 & 28.78 & 46.69 & 42.41 & 39.71 \\
& NAS~\cite{yeo2018neural} & 36.27 & 32.08 & 29.94 & 40.70 & 34.01 & 28.84 & 46.78 & 42.53 & 39.76 \\ 
& CaFM\cite{liu2021overfitting} & 36.63 & 32.21 & 29.98 & 40.76 & 34.08 & 29.93 & 46.98 & 42.62 & 39.81 \\
& EMT~\cite{li2022efficient} & 36.70 & 32.27 & 29.94 & 40.79 & 34.15 & 29.99 & 47.06 & 42.72 & 39.82 \\
& \cellcolor{green!10}\bf{RepCaM++} & \cellcolor{green!10}\textbf{36.93} & \cellcolor{green!10}\textbf{32.61} & \cellcolor{green!10}\textbf{30.13} & \cellcolor{green!10}\textbf{40.94} & \cellcolor{green!10}\textbf{34.41} & \cellcolor{green!10}\textbf{30.12} & \cellcolor{green!10}\textbf{47.38} & \cellcolor{green!10}\textbf{42.92} & \cellcolor{green!10}\textbf{39.98} \\
\midrule 
\multirow{5}{*}{VDSR}& awDNN~\cite{yeo2017will} & 41.27 & 35.03 & 32.16 & 44.16 & 35.99 & 30.65 & 48.18 & 43.03 & 41.07\\
& NAS~\cite{yeo2018neural} & 42.53 & 35.97 & 33.86 &  44.71  & 36.57 & 31.05 & 48.49 &43.41 & 41.33\\
& CaFM\cite{liu2021overfitting} & 43.02 & 36.17 & 33.98 & 44.85 & 36.46 & 31.08 & 48.61 & 43.62 & 41.49\\
& EMT~\cite{li2022efficient} & 43.11 & 36.23 & 34.07 & 44.93 & 36.44 & 31.09 & 48.75 & 43.65 & 41.57\\
& \cellcolor{green!10}\bf{RepCaM++} & \cellcolor{green!10}\textbf{43.32} & \cellcolor{green!10}\textbf{36.41} & \cellcolor{green!10}\textbf{34.35} & \cellcolor{green!10}\textbf{45.04} & \cellcolor{green!10}\textbf{36.57} & \cellcolor{green!10}\textbf{31.34} & \cellcolor{green!10}\textbf{48.91} & \cellcolor{green!10}\textbf{43.89} & \cellcolor{green!10}\textbf{41.90} \\
\midrule
\multirow{5}{*}{EDSR} &   awDNN~\cite{yeo2017will} & 42.24 & 35.88 & 33.44 & 43.06 & 37.89 & 34.94 & 48.87 & 44.51 & 42.58\\ 
& NAS~\cite{yeo2018neural} & 42.82 & 36.42 & 34.00 & 45.06 & 38.38 & 35.47 & 49.10 & 44.80 & 42.83 \\
&CaFM~\cite{liu2021overfitting} & 43.13 & 37.04 & 34.47 & 45.35 & 38.66 & 35.70 & 49.30 & 45.03 & 43.12  \\
&EMT~\cite{li2022efficient} & 43.04 & 37.14 & 34.54 & 45.42 & 38.72 & 35.83 & 49.25 & 45.11 & 43.22  \\
& \cellcolor{green!10}\bf{RepCaM++} & \cellcolor{green!10}\textbf{43.32} & \cellcolor{green!10}\textbf{37.23} & \cellcolor{green!10}\textbf{34.76} & \cellcolor{green!10}\textbf{45.72} & \cellcolor{green!10}\textbf{38.91} & \cellcolor{green!10}\textbf{36.00} & \cellcolor{green!10}\textbf{49.45} & \cellcolor{green!10}\textbf{45.37} & \cellcolor{green!10}\textbf{43.46} \\ 

\midrule
\midrule

&  & \multicolumn{3}{c}{\bf { game-45s-PSNR $\uparrow$ }}& \multicolumn{3}{c}{\bf { inter-45s-PSNR $\uparrow$ }} & \multicolumn{3}{c}{\bf {vlog-45s-PSNR $\uparrow$}}  \\
\bf{Model}& \bf{Method} & $\times$2 & $\times$3 & $\times$4 & $\times$2 & $\times$3 & $\times$4 & $\times$2 & $\times$3 & $\times$4 \\
\midrule 
\multirow{5}{*}{ESPCN} & awDNN~\cite{yeo2017will} & 35.42 & 30.63 & 28.65 & 38.64 & 31.97 & 28.32 & 45.71 & 41.40 & 39.20 \\
& NAS~\cite{yeo2018neural} & 35.55 & 30.67 & 28.74 & 38.81 & 32.14 & 28.61 & 45.81 & 41.52 & 39.29 \\
& CaFM\cite{liu2021overfitting} & 36.09 & 31.06 & 29.05 & 38.88 & 32.22 & 28.75 & 46.19 & 41.72 & 39.52  \\
& EMT~\cite{li2022efficient} & 36.63 & 31.33 & 29.56 & 39.00 & 32.45 & 29.11 & 46.25 & 41.81 & 39.66  \\
& \cellcolor{green!10}\bf{RepCaM++} & \cellcolor{green!10}\textbf{37.42} & \cellcolor{green!10}\textbf{31.86} & \cellcolor{green!10}\textbf{29.69} & \cellcolor{green!10}\textbf{39.52} & \cellcolor{green!10}\textbf{32.69} & \cellcolor{green!10}\textbf{29.20} & \cellcolor{green!10}\textbf{46.44} & \cellcolor{green!10}\textbf{41.90} & \cellcolor{green!10}\textbf{39.80}  \\
\midrule
\multirow{5}{*}{SRCNN} & awDNN~\cite{yeo2017will} & 35.05 & 30.50 & 28.59 & 38.66 & 31.78 & 28.25 & 45.87 & 41.58 &39.29 \\
& NAS~\cite{yeo2018neural} & 35.15 & 30.55 & 28.61 & 38.79 & 31.93 & 28.38 & 45.95 & 41.66 & 39.36 \\ 
& CaFM\cite{liu2021overfitting} & 35.49 & 30.63 & 28.66 & 38.88 & 32.02 & 28.48 & 46.18 & 41.85 & 39.52 \\
& EMT~\cite{li2022efficient} & 35.61 & 30.77 & 28.83 & 38.90 & 32.22 & 28.71 & 46.23 & 41.98 & 39.58 \\
& \cellcolor{green!10}\bf{RepCaM++} & \cellcolor{green!10}\textbf{35.77} & \cellcolor{green!10}\textbf{30.94} & \cellcolor{green!10}\textbf{28.95} & \cellcolor{green!10}\textbf{39.04} & \cellcolor{green!10}\textbf{32.45} & \cellcolor{green!10}\textbf{28.97} & \cellcolor{green!10}\textbf{46.53} & \cellcolor{green!10}\textbf{42.13} & \cellcolor{green!10}\textbf{39.78} \\
\midrule 
\multirow{5}{*}{VDSR}& awDNN~\cite{yeo2017will } & 40.29 & 34.53 & 31.28 & 41.99 & 33.80 & 30.34 & 47.61 & 42.92 & 40.94 \\
& NAS~\cite{yeo2018neural} & 41.37 & 34.92 & 32.42 & 42.40 & 34.53 & 31.10 & 47.88 & 43.33 & 41.23\\
& CaFM\cite{liu2021overfitting}  & 41.92 & 35.56 & 33.16 & 42.86 & 34.49 & 30.95 & 48.00 & 43.50 & 41.38\\
& EMT~\cite{li2022efficient} & 42.07 & 35.87 & 33.34 & 42.96 & 34.63 & 31.10 & 48.21 & 43.67 & 41.44\\
& \cellcolor{green!10}\bf{RepCaM++} & \cellcolor{green!10}\textbf{42.31} & \cellcolor{green!10}\textbf{36.04} & \cellcolor{green!10}\textbf{33.51} & \cellcolor{green!10}\textbf{43.22} & \cellcolor{green!10}\textbf{34.94} & \cellcolor{green!10}\textbf{31.34} & \cellcolor{green!10}\textbf{48.46} & \cellcolor{green!10}\textbf{43.97} & \cellcolor{green!10}\textbf{41.56}  \\
\midrule
\multirow{5}{*}{EDSR}& awDNN~\cite{yeo2017will}  & 42.11 & 35.75 & 33.33 & 42.73 & 34.49 & 31.34 & 47.98 & 43.58 & 41.53\\ 
& NAS~\cite{yeo2018neural} & 43.22 & 36.72 & 34.32 & 43.31 & 35.80 & 32.67 & 48.48 & 44.12 & 42.12 \\
& CaFM\cite{liu2021overfitting} & 43.32 & 37.19 & 34.61 & 43.37 & 35.62 & 32.35 & 48.45 & 44.11 & 42.16\\
& EMT~\cite{li2022efficient} & 43.35 & 37.54 & 34.77 & 43.41 & 35.76 & 32.76 & 48.41 & 44.23 & \textbf{42.25}\\
& \cellcolor{green!10}\bf{RepCaM++} & \cellcolor{green!10}\textbf{43.61} & \cellcolor{green!10}\textbf{37.84} & \cellcolor{green!10}\textbf{35.09} & \cellcolor{green!10}\textbf{43.47} & \cellcolor{green!10}\textbf{36.01} & \cellcolor{green!10}\textbf{33.74} & \cellcolor{green!10}\textbf{48.48} & \cellcolor{green!10}\textbf{44.38} & \cellcolor{green!10}42.24 \\ 
\bottomrule
\end{tabular}}
\label{tab:main_psnr}
\vspace{-0.3cm}
\end{table*}

\subsection{Comparison with other methods}
\label{sec:main}

In this section, we compare our method against other neural video delivery methods \cite{yeo2017will,yeo2018neural,liu2021overfitting,li2022efficient} with various super-resolution backbones including \cite{lim2017enhanced, shi2016real, kim2016accurate, dong2015image}.

\noindent\textbf{Restoration performance.} In this section, we compare our method with state-of-the-art techniques that utilize either general model overfitting or time-divided model overfitting on different SR backbones. Due to space constraints, we sample three video categories from VSD4K and test on two different video durations, including 15 seconds and 45 seconds. As shown in \cref{tab:long}, RepCaM++ consistently outperforms the CaFM and EMT methods in video super-resolution tasks across various scaling factors and different backbone models, underscoring its strong generalization capability. For instance, RepCaM++ achieves a PSNR of 38.44 dB for $\times$2, 33.51 dB for $\times$3, and 31.31 dB for $\times$4 when using the ESPCN backbone, outperforming both CaFM and EMT on Game-15s. Similarly, with the SRCNN backbone in the Inter-15s, RepCaM++ attains PSNR values of 40.94 dB, 34.41 dB, and 30.12 dB for the respective scaling factors, again surpassing the baseline methods. Furthermore, when evaluated on the Vlog-15 with the EDSR backbone, RepCaM++ achieves PSNR values of 49.45 dB for $\times$2, 45.37 dB for $\times$3, and 43.46 dB for $\times$4, maintaining its superiority. These results highlight the robustness and adaptability of RepCaM++ in handling different scaling and backbone variations.

To further demonstrate the effectiveness of our proposed methods from various perspectives, we provide additional experiment results with LPIPS and SSIM conducted on the left three VSD4K datasets as shown in \cref{tab:lpips}. Notably, RepCaM++ demonstrates a significant advantage over other methods across all scales and scenarios. In terms of LPIPS, RepCaM++ achieves the lowest scores in all categories, with values as low as 0.0032 for the EDSR backbone at scale $\times$2 on the city dataset. This suggests that RepCaM++ produces results that are much closer to the ground truth compared to its counterparts. Similarly, RepCaM++ outperforms all other methods in SSIM, with scores reaching as high as 0.9981 for the EDSR backbone at scale $\times$2 on the city dataset. These results emphasize the robustness and effectiveness of RepCaM++ in enhancing visual quality in NVD. As RepCaM++ enhances performance by integrating multiple feature extraction across different scales with TVPs, effectively capturing a wider range of features from fine details to global structures, while the multi-branch architecture enables simultaneous learning of texture and structural information, improving generalization to new data and enhancing robustness to variations in input.

Last but not least, we delved into video smoothness by evaluating the Mean Square Error (MSE) ($\times10^{-1}$) between LR and downsampled SR videos following~\cite{gao2023implicit}. We compiled the average consistency results in \cref{tab:par}, spanning six scenes from the VSD4K dataset. The findings indicate that baseline models exhibit poor consistency values, suggesting a significant deviation of their SR outputs from the original LR images. In contrast, RepCaM++ surpasses the state-of-the-art EMT by 0.22, demonstrating superior video smoothness and frame consistency, thanks to the integrated unified RepCaM framework and the detail-enhancing TVP.

\noindent\textbf{Delivery size.} To better show the effectiveness of RepCaM, we compare the delivery size with other baselines, including two end-to-end DNN-based video compression methods SRVC~\cite{khani2021efficient} and DVC~\cite{lu2019dvc} in \cref{fig:size}. Without bringing extra computational costs, RepCaM (indicated as 'Ours' in \cref{fig:size}.) can reduce delivery costs and improve the quality of video restoration. Meanwhile, as shown in \cref{tab:par}, RepCaM++, which has almost the same model size as RepCaM, maintains the same HR video size as other methods while achieving the smallest combined size for LR video and model (LR+Model) at just 3.89 MB, indicating superior transmission and storage efficiency. Despite the smallest model size, our method achieves a PSNR of 33.74 dB, surpassing other methods and demonstrating higher compression efficiency without compromising video quality. Additionally, our method matches the computational complexity of others, with FLOPS of 9.18G, achieving superior performance without added computational overhead. Since the size of TVPs, termed as $T$, is extremely limited compared with other transmission costs, the overall transmission costs can be approximated as $S+N\times L$. While RepCaM++ incurs elevated GPU memory consumption and extended training duration, it is critical to emphasize that these requirements are exclusively confined to the training phase conducted on computationally resource-abundant central servers. Compared with previous baselines, RepCaM++ scales up the SR model size with multi-branch architecture during training to enhance the model overfitting ability, thus leading to better content awareness of the SR backbone.

\begin{table*}[t]
\linespread{0.9}
\small
\centering
\caption{LPIPS and SSIM results of RepCaM++ on sport, dance, and city VSD4K data with different SR backbones.}
\scalebox{0.925}{
\begin{tabular}{l|l|p{1.3cm}p{1.3cm}p{1.3cm}|p{1.3cm}p{1.3cm}p{1.3cm}|p{1.3cm}p{1.3cm}p{1.3cm}}
\toprule
&  & \multicolumn{3}{c}{\bf { sport-45s-LPIPS $\downarrow$ }}& \multicolumn{3}{c}{\bf { dance-45s-LPIPS $\downarrow$ }} & \multicolumn{3}{c}{\bf {city-45s-LPIPS $\downarrow$}}  \\
\bf { Model } & \bf { Method } & $\times$2 & $\times$3 & $\times$4 & $\times$2 & $\times$3 & $\times$4 & $\times$2 & $\times$3 & $\times$4 \\
\midrule 
\multirow{5}{*}{ESPCN} & awDNN~\cite{yeo2017will} & 0.0139 & 0.0313 & 0.0544 & 0.0152 & 0.0378 & 0.0601 & 0.0249 & 0.0582 & 0.1191  \\
& NAS~\cite{yeo2018neural} & 0.0131 & 0.0306 & 0.0537 & 0.0141 & 0.0367 & 0.0598 & 0.0244 & 0.0578 & 0.1171 \\
& CaFM\cite{liu2021overfitting} & 0.0116 & 0.0299 & 0.0517 & 0.0125 & 0.0342 & 0.0589 & 0.0231 & 0.0556 & 0.1152 \\
& EMT~\cite{li2022efficient}& 0.0097 & 0.0287 & 0.0497 & 0.0101 & 0.0313 & 0.0546 & 0.0198 & 0.0521 & 0.1133 \\
& \cellcolor{green!10}\bf{RepCaM++} & \cellcolor{green!10}\textbf{0.0048} & \cellcolor{green!10}\textbf{0.0198} & \cellcolor{green!10}\textbf{0.0426} & \cellcolor{green!10}\textbf{0.0058} & \cellcolor{green!10}\textbf{0.0253} & \cellcolor{green!10}\textbf{0.0473} & \cellcolor{green!10}\textbf{0.0115} & \cellcolor{green!10}\textbf{0.0484} & \cellcolor{green!10}\textbf{0.1055} \\
\midrule
\multirow{5}{*}{EDSR} &   awDNN~\cite{yeo2017will} & 0.0098 & 0.0299 & 0.0423 & 0.0101 & 0.0296 & 0.0400 & 0.0115 & 0.0377 & 0.0590\\ 
& NAS~\cite{yeo2018neural} & 0.0098 & 0.0278 & 0.0411 & 0.0095 & 0.0275 & 0.0389 & 0.0107 & 0.0353 & 0.0585  \\
&CaFM~\cite{liu2021overfitting} & 0.0085 & 0.0241 & 0.0388 & 0.0089 & 0.0256 & 0.0364 & 0.0093 & 0.0301 & 0.0543  \\
&EMT~\cite{li2022efficient} & 0.0063 & 0.0241 & 0.0331 & 0.0079 & 0.0217 & 0.0306 & 0.0088 & 0.0277 & 0.0503  \\
& \cellcolor{green!10}\bf{RepCaM++} & \cellcolor{green!10}\textbf{0.0032} & \cellcolor{green!10}\textbf{0.0133} & \cellcolor{green!10}\textbf{0.0276} & \cellcolor{green!10}\textbf{0.0037} & \cellcolor{green!10}\textbf{0.0155} & \cellcolor{green!10}\textbf{0.0275} & \cellcolor{green!10}\textbf{0.0049} & \cellcolor{green!10}\textbf{0.0208} & \cellcolor{green!10}\textbf{0.0437} \\ 

\midrule
\midrule

&  & \multicolumn{3}{c}{\bf { sport-45s-SSIM $\uparrow$ }}& \multicolumn{3}{c}{\bf { dance-45s-SSIM $\uparrow$ }} & \multicolumn{3}{c}{\bf {city-45s-SSIM $\uparrow$}}  \\
\bf{Model}& \bf{Method} & $\times$2 & $\times$3 & $\times$4 & $\times$2 & $\times$3 & $\times$4 & $\times$2 & $\times$3 & $\times$4 \\
\midrule 
\multirow{5}{*}{ESPCN} & awDNN~\cite{yeo2017will} & 0.9931 & 0.9835 & 0.9734 & 0.9857 & 0.9713 & 0.9626 & 0.9886 & 0.9580 & 0.9373 \\
& NAS~\cite{yeo2018neural} & 0.9933 & 0.9841 & 0.9747 & 0.9876 & 0.9724 & 0.9634 & 0.9891 & 0.9584 & 0.9381 \\
& CaFM\cite{liu2021overfitting} & 0.9941 & 0.9858 & 0.9767 & 0.9895 & 0.9734 & 0.9656 & 0.9903 & 0.9599 & 0.9387  \\
& EMT~\cite{li2022efficient} & 0.9952 & 0.9887 & 0.9785 & 0.9903 & 0.9793 & 0.9689 & 0.9910 & 0.9626 & 0.9421  \\
& \cellcolor{green!10}\bf{RepCaM++} & \cellcolor{green!10}\textbf{0.9973} & \cellcolor{green!10}\textbf{0.9917} & \cellcolor{green!10}\textbf{0.9837} & \cellcolor{green!10}\textbf{0.9962} & \cellcolor{green!10}\textbf{0.9833} & \cellcolor{green!10}\textbf{0.9760} & \cellcolor{green!10}\textbf{0.9924} & \cellcolor{green!10}\textbf{0.9750} & \cellcolor{green!10}\textbf{0.9583}  \\
\midrule 
\multirow{5}{*}{EDSR}& awDNN~\cite{yeo2017will}  & 0.9954 & 0.9917 & 0.9856 & 0.9937 & 0.9870 & 0.9802 & 0.9885 & 0.9811 & 0.9714\\ 
& NAS~\cite{yeo2018neural} & 0.9956 & 0.9921 & 0.9861 & 0.9944 & 0.9875 & 0.9808 & 0.9893 & 0.9821 & 0.9735 \\
& CaFM\cite{liu2021overfitting} & 0.9961 & 0.9926 & 0.9871 & 0.9951 & 0.9878 & 0.9813 & 0.9901 & 0.9823 & 0.9771\\
& EMT~\cite{li2022efficient} & 0.9968 & 0.9931 & 0.9882 & 0.9956 & 0.9887 & 0.9821 & 0.9923 & 0.9837 & 0.9799\\
& \cellcolor{green!10}\bf{RepCaM++} & \cellcolor{green!10}\textbf{0.9981} & \cellcolor{green!10}\textbf{0.9947} & \cellcolor{green!10}\textbf{0.9902} & \cellcolor{green!10}\textbf{0.9975} & \cellcolor{green!10}\textbf{0.9903} & \cellcolor{green!10}\textbf{0.9846} & \cellcolor{green!10}\textbf{0.9967} & \cellcolor{green!10}\textbf{0.9891} & \cellcolor{green!10}\textbf{0.9818} \\ 
\bottomrule
\end{tabular}}
\label{tab:lpips}
\end{table*}



\begin{table*}[t]
      \begin{center}
        \caption{Comparisons with other methods on all the videos in VSD4K-45s. The FLOPS(G) is calculated on LR (270$\times$480). The training time is reported as X seconds per iteration (batch size=256).}
        \resizebox{0.975\textwidth}{!}{
        \setlength{\tabcolsep}{1mm}{
       	\begin{tabular}{c|cccccc}
       	\toprule
     Model & Communication Cost (MB) $\downarrow$ & PSNR $\uparrow$ & Consistency $\downarrow$ & FLOPS(G) $\downarrow$ & GPU Memory (MB) $\downarrow$& Training Time $\downarrow$\\

        \midrule
     NAS~\cite{yeo2018neural}  & 3.62+2.43 (6.05) & 32.87 & 4.083 & 9.18 & \textbf{2341} & \textbf{0.0014}\\
     CaFM\cite{liu2021overfitting}  & 3.62+0.29 (3.91) & 33.12 & 3.707 & 9.45 & 2688 & 0.0021\\
     EMT\cite{li2022efficient}  & 3.62+0.34 (3.96) & 33.36 & 3.542 & 9.18 & 3123 & 0.0027\\
     \cellcolor{green!10}\textbf{RepCaM++}  & \cellcolor{green!10}\textbf{3.62+0.27 (3.89)} & \cellcolor{green!10}\textbf{33.74} & \cellcolor{green!10}\textbf{3.320} & \cellcolor{green!10}\textbf{9.18} & \cellcolor{green!10}6421 & \cellcolor{green!10}0.0046 \\
        \bottomrule
		\end{tabular}}
  }
      \label{tab:par}
      \end{center}
      \vspace{-0.2cm}
\end{table*}

\noindent\textbf{Extension to long videos.} We also extend RepCaM++ to long videos beyond 1 minute, and the results are shown in \cref{tab:long}. We can observe that RepCaM++ exhibits superior performance compared to CaFM and EMT when applied to long video sequences, demonstrating its ability to better handle extended durations and scene variability. For example, RepCaM++ achieves a PSNR of 43.81 dB for $\times$2, 37.41 dB for $\times$3, and 34.77 dB for $\times$4, surpassing both CaFM and EMT on game-1min. Additionally, RepCaM++ maintains its lead with a PSNR of 43.33 dB for $\times$2 and 34.45 dB for $\times$4 on even longer game-5min. These results highlight the effectiveness of our method, which integrates TVPs that focus on overfitting specific video chunks to enhance video comprehension and adaptability to scene change.

\begin{table}[t]
      \begin{center}
        \caption{Ablation study on Transparent Visual Prompt with different number and size on $\times$4.}
        \resizebox{0.475\textwidth}{!}{
        \setlength{\tabcolsep}{3mm}{
       	\begin{tabular}{c|cccc}
       	\toprule
     Settings & game-45s & inter-45s & vlog-45s & sport-45s \\

        \midrule
     \cellcolor{green!10}9 TVP& \cellcolor{green!10}\textbf{35.09} & \cellcolor{green!10}\textbf{33.73} & \cellcolor{green!10}\textbf{42.23} & \cellcolor{green!10}\textbf{40.26} \\
     27 TVP & 35.01 & 32.85 & 42.21 & 40.15\\
     \midrule
     32 $\times$ 32  & 35.02 & \textbf{33.75} & 42.25 & 40.25\\
     \cellcolor{green!10}48 $\times$ 48 & \cellcolor{green!10}\textbf{35.09} & \cellcolor{green!10}33.73 & \cellcolor{green!10}\textbf{42.23} & \cellcolor{green!10}40.26 \\
     64 $\times$ 64 & 25.12 & 33.71 & 32.22& \textbf{40.29}\\
        \bottomrule
		\end{tabular}}
  }
      \label{tvp}
      \end{center}
\end{table}

\begin{figure}[t]
\includegraphics[width=0.5\textwidth]{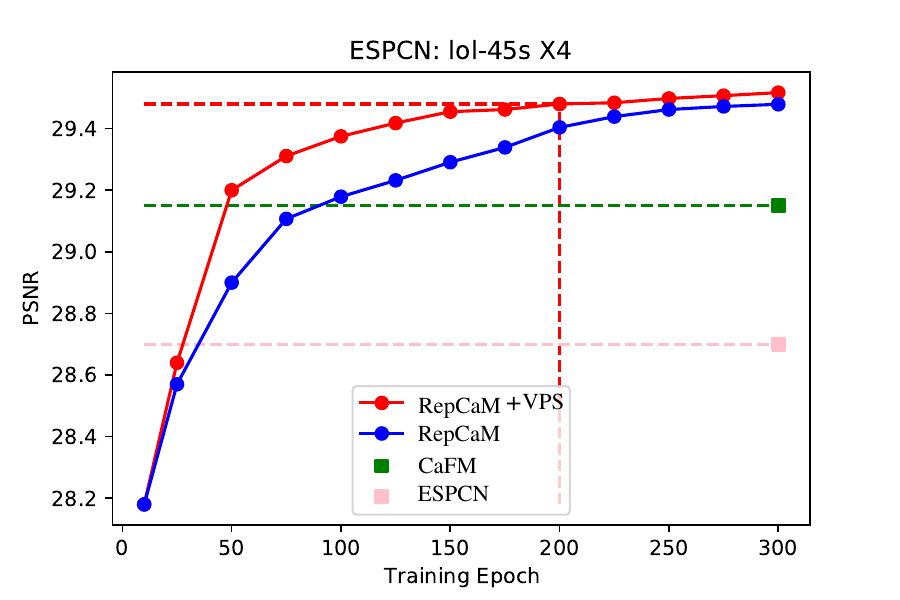}
\centering
\caption{Results on the effectiveness of VPS. The X-axis is the number of training epochs, and the Y-axis is the PSNR.}
\vspace{-0.4cm}
\label{vps}
\end{figure}

\subsection{Ablation study}
\label{sec:abla}
\textbf{Variants of the transparent visual prompt.} 
We explore the performance of the SR model with different variants of TVPs, as shown in \cref{tvp}. TVPs are lightweight, learnable modules inserted into intermediate feature maps to adaptively refine spatial-temporal features for super-resolution tasks. However, as indicated by the upper part of \cref{tvp}, increasing the number of TVPs (e.g., from 9 to 27) leads to diminishing returns, as overlapping receptive fields introduce redundancy and suppress critical high-frequency details (e.g., edges, textures). This highlights the importance of balancing the number of TVPs to capture both unique chunk characteristics and shared commonalities. Additionally, the lower part of \cref{tvp} shows that different TVP sizes (e.g., 32 × 32, 48 × 48, 64 × 64) yield optimal performance for different video categories. Since TVPs often focus on foreground samples, these results emphasize the need to balance the learning of dynamic foreground and background content. Overall, TVPs enhance the model's perception of continuously changing foreground objects rather than merely increasing overfitting through TVPs.

\begin{table*}[t]
\linespread{0.9}
\small
\centering
\caption{Comparison results of RepCaM++ with different data overfitting methods on long videos.}
\scalebox{0.925}{
\begin{tabular}{l|l|p{1.3cm}p{1.3cm}p{1.3cm}|p{1.3cm}p{1.3cm}p{1.3cm}|p{1.3cm}p{1.3cm}p{1.3cm}}
\toprule
&  & \multicolumn{3}{c}{\bf { game-1min }}& \multicolumn{3}{c}{\bf { game-2min }} & \multicolumn{3}{c}{\bf {game-5min}}  \\
\bf { Model } & \bf { Method } & $\times$2 & $\times$3 & $\times$4 & $\times$2 & $\times$3 & $\times$4 & $\times$2 & $\times$3 & $\times$4 \\
\midrule 
\multirow{5}{*}{EDSR} &   awDNN~\cite{yeo2017will} & 41.82 & 35.25 & 32.61 & 41.89 & 35.72 & 33.27 & 40.62 & 34.59 & 32.14\\ 
& NAS~\cite{yeo2018neural} & 43.24 & 36.56 & 33.52 & 43.20 & 37.00 & 34.47 & 42.47 & 36.08 & 33.53 \\
&CaFM~\cite{liu2021overfitting} & 43.49 & 37.18 & 34.33 & 43.49 & 37.47 & 34.80 & 43.01 & 36.65 & 34.07 \\
&EMT~\cite{li2022efficient} & 43.61 & 37.26 & 34.45 & 43.61 & 37.58 & 34.91 & 43.18 & \textbf{36.77} & 34.20 \\
& \cellcolor{green!10}RepCaM & \cellcolor{green!10}43.72 & \cellcolor{green!10}37.33 & \cellcolor{green!10}34.58 & \cellcolor{green!10}43.66 & \cellcolor{green!10}37.67 & \cellcolor{green!10}34.99 & \cellcolor{green!10}43.25 & \cellcolor{green!10}36.66 & \cellcolor{green!10}34.36 \\ 
& \cellcolor{green!10}\bf{RepCaM++} & \cellcolor{green!10}\textbf{43.81} & \cellcolor{green!10}\textbf{37.41} & \cellcolor{green!10}\textbf{34.77} & \cellcolor{green!10}\textbf{43.72} & \cellcolor{green!10}\textbf{37.88} & \cellcolor{green!10}\textbf{35.12} & \cellcolor{green!10}\textbf{43.33} & \cellcolor{green!10}36.65 & \cellcolor{green!10}\textbf{34.45} \\ 

\bottomrule
\end{tabular}}
\label{tab:long}
\end{table*}

\begin{table*}[t]
\linespread{0.9}
\small
\centering
\caption{Performance comparisons with H.264/H.265 and CaFM based on EDSR backbone with 45s videos.}
\scalebox{0.925}{
\begin{tabular}{l|l|p{1.3cm}p{1.3cm}p{1.3cm}|p{1.3cm}p{1.3cm}p{1.3cm}|p{1.3cm}p{1.3cm}p{1.3cm}}
\toprule
&  & \multicolumn{3}{c}{\bf { game-45s }}& \multicolumn{3}{c}{\bf { vlog-45s }} & \multicolumn{3}{c}{\bf {inter-45s}}  \\
\bf { Model } & \bf { Method } & $\times$2 & $\times$3 & $\times$4 & $\times$2 & $\times$3 & $\times$4 & $\times$2 & $\times$3 & $\times$4 \\
\midrule 
\multirow{2}{*}{Codecs} & H.264 & 39.26& 36.99& 35.52& 43.45& 42.07& 41.31 &37.99& 36.07 &35.06 \\
& H.265 & 39.77 &37.71& 36.42& 44.24& 43.09& 42.31 &38.31 &36.51 &35.58 \\
\midrule
\multirow{3}{*}{EDSR} & CaFM & 43.32 & 37.19 & 34.61 & 48.45 & 44.11 & 42.16 & 43.37 & 35.62 & 32.35 \\
& \cellcolor{green!10}RepCaM & \cellcolor{green!10}43.46 & \cellcolor{green!10}37.78 & \cellcolor{green!10}34.88 & \cellcolor{green!10}48.46 & \cellcolor{green!10}44.22 & \cellcolor{green!10}42.21 & \cellcolor{green!10}41.40 & \cellcolor{green!10}35.83 & \cellcolor{green!10}33.15 \\
& \cellcolor{green!10}\bf{RepCaM++} & \cellcolor{green!10}\textbf{43.61} & \cellcolor{green!10}\textbf{37.84} & \cellcolor{green!10}\textbf{35.09} & \cellcolor{green!10}\textbf{48.48} & \cellcolor{green!10}\textbf{44.38} & \cellcolor{green!10}\textbf{42.24} & \cellcolor{green!10}\textbf{43.47} & \cellcolor{green!10}\textbf{36.01} & \cellcolor{green!10}\textbf{33.74} \\

\midrule
\midrule

& & \multicolumn{3}{c}{\bf { sport-45s }}& \multicolumn{3}{c}{\bf { dance-45s }} & \multicolumn{3}{c}{\bf {city-45s}}  \\
\bf { Model } & \bf { Method } & $\times$2 & $\times$3 & $\times$4 & $\times$2 & $\times$3 & $\times$4 & $\times$2 & $\times$3 & $\times$4 \\
\midrule 
\multirow{2}{*}{Codecs} & H.264 & 40.30 &38.09 &36.83 &31.11 &28.32 &26.76 &36.60& 34.18 &32.89 \\
& H.265 & 41.35 &39.66 &38.67 &32.62& 30.40 &29.18 &37.17 &35.10& 34.03 \\
\midrule
\multirow{3}{*}{EDSR} & CaFM & 48.24 & 42.93 & 40.21 & \textbf{45.97} & \textbf{39.96} & 37.84 & 39.75 & 34.48 & 31.87 \\ 
& \cellcolor{green!10}RepCaM & \cellcolor{green!10}48.31 & \cellcolor{green!10}43.04 & \cellcolor{green!10}40.23 & \cellcolor{green!10}45.57 & \cellcolor{green!10}39.41 & \cellcolor{green!10}37.87 & \cellcolor{green!10}39.77 & \cellcolor{green!10}34.50 & \cellcolor{green!10}31.89 \\
& \cellcolor{green!10}\bf{RepCaM++} & \cellcolor{green!10}\textbf{48.59} & \cellcolor{green!10}\textbf{43.27} & \cellcolor{green!10}\textbf{40.27} & \cellcolor{green!10}45.86 & \cellcolor{green!10}39.67 & \cellcolor{green!10}\textbf{38.28} & \cellcolor{green!10}\textbf{39.81} & \cellcolor{green!10}\textbf{34.55} & \cellcolor{green!10}\textbf{31.91} \\
\bottomrule
\end{tabular}}
\label{tab:h264}
\end{table*}

\noindent\textbf{Variants of the number of branches.}
Since we implement the parallel-cascade RepCaM architecture as a three-branch network, we also study the effect of the number of branches. The ablation study presented in \cref{branch} evaluates the impact of varying the number of parallel branches in the RepCaM module using EDSR as the backbone on 45-second video datasets. The results indicate that increasing the number of branches generally enhances performance across all datasets. For example, performance improves from 42.55 dB with a 1-branch configuration to 43.12 dB with 2-branches and further to 43.61 dB with 3 branches on game-45s. The gain continues with 4 and 5-branches, reaching 43.88 dB and 43.96 dB, respectively. Similar trends are observed in other scenarios. Thus, we confirm that the experimental results align with our observation in the \cref{motivation} that each branch provides a different level of feature perception, leading to a more comprehensive data overfitting. Moreover, while additional branches consistently improve performance, the most substantial gains are realized with the 3-branch configuration, indicating a trade-off between computational costs and performance enhancement can be made under different conditions.

\begin{figure*}[t]
\includegraphics[width=0.98\textwidth]{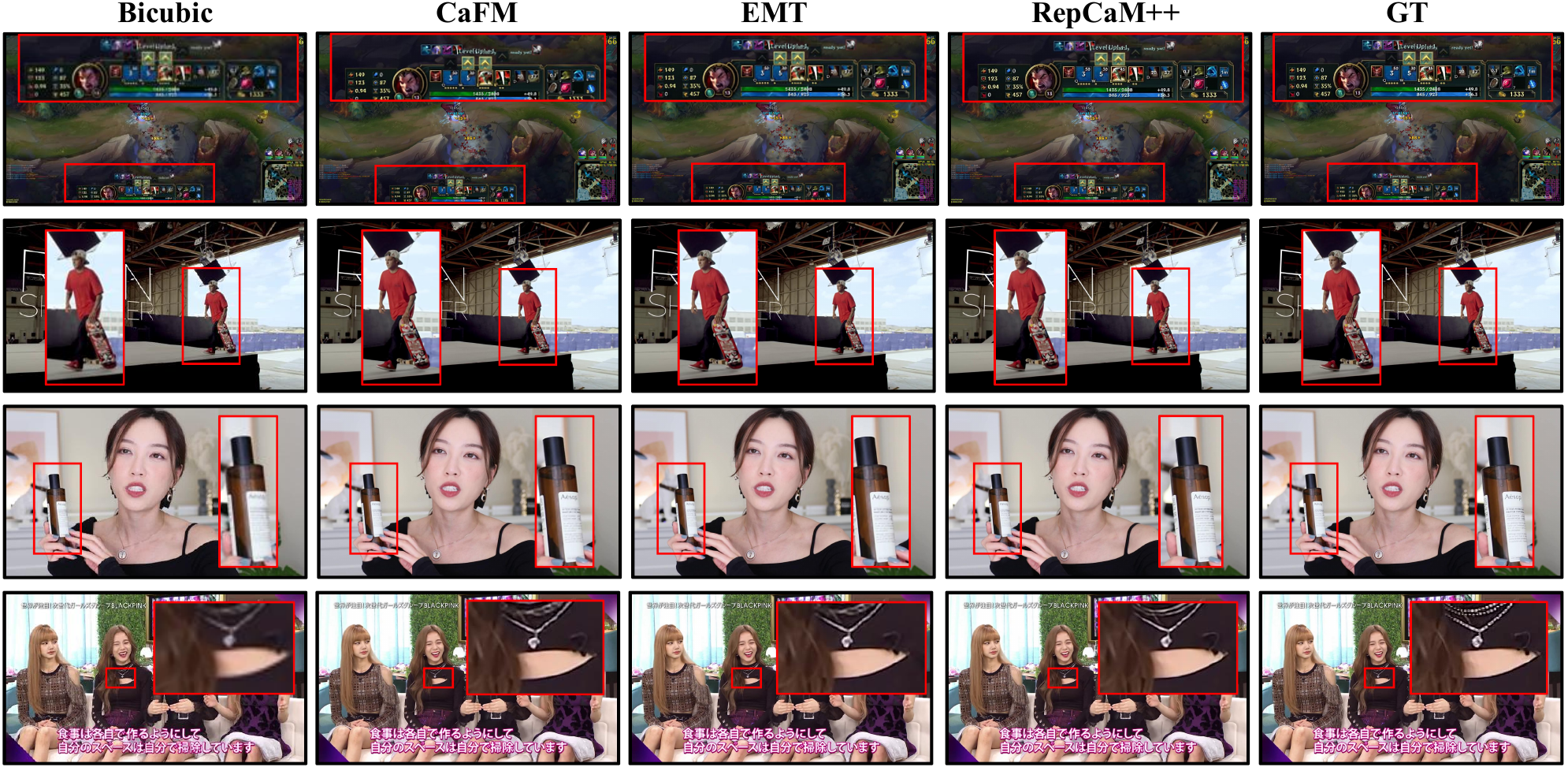}
\centering
\caption{Qualitative comparisons on the VSD4K dataset across the game, sport, vlog, and inter scenes. Best viewed by zooming x4.}
\label{fig:vis}
\end{figure*}

\begin{figure}[t]
\includegraphics[width=0.48\textwidth]{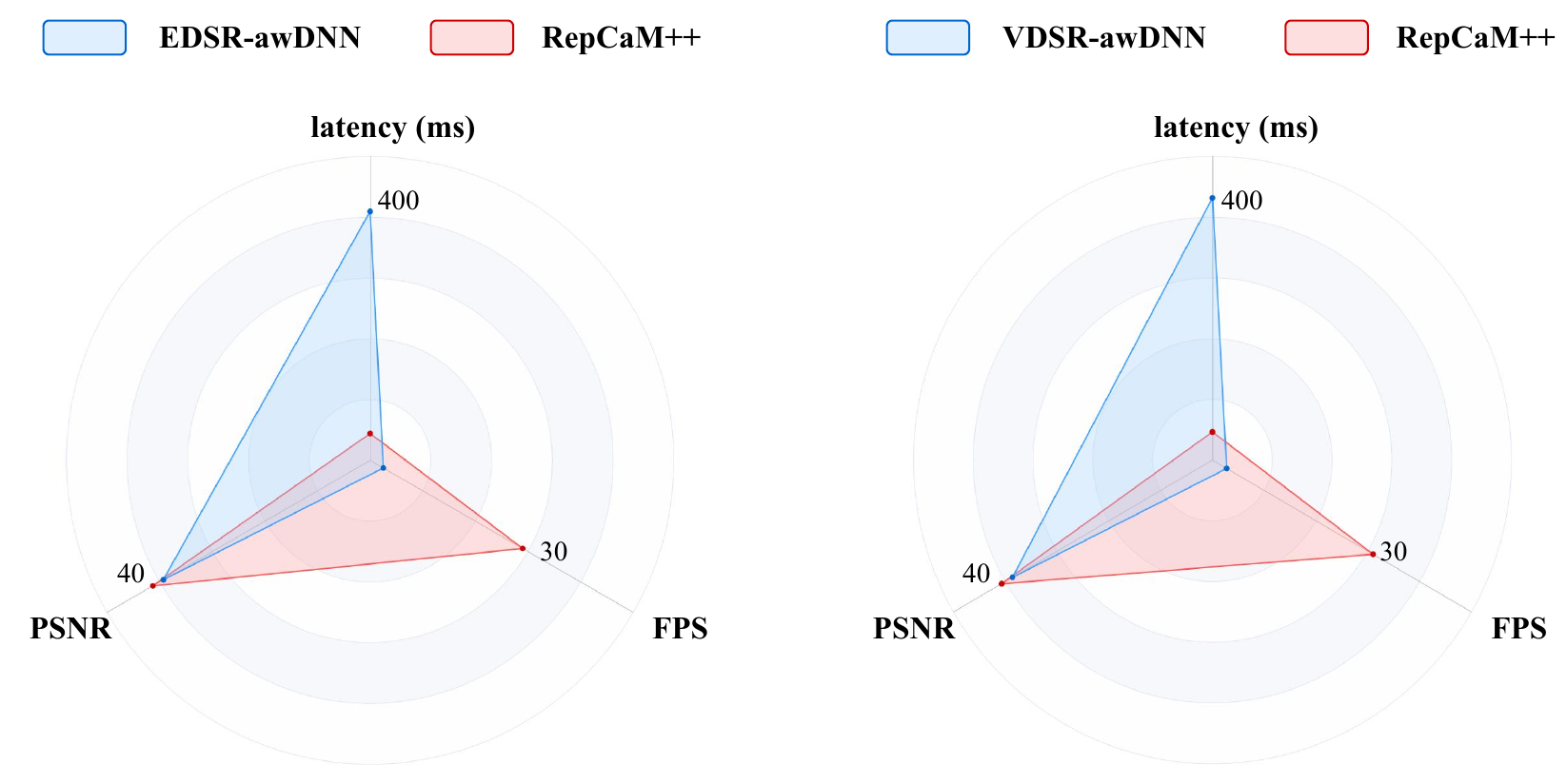}
\centering
\caption{Execution speed and video quality comparison between RepCaM++ and EDSR, VDSR with a Huawei phone.}
\label{fig:app}
\vspace{-0.5cm}
\end{figure}

\noindent\textbf{Benefit of VPS.} To evaluate the effectiveness of VPS, we compared the PSNR of the RepCaM model with and without VPS across different training epochs, as shown in \cref{vps}. ESPCN recommends its direct use for NVD. Our findings demonstrate that VPS significantly accelerates convergence relative to other baselines. Additionally, the performance results show that VPS allows the RepCaM model to adapt and modulate more precisely by selectively enhancing its perception of informative samples in temporally redundant video sequences.

\noindent\textbf{Effectiveness of RepCaM and TVP.}
The comparative analysis in \cref{tab:long} highlights the performance improvements of RepCaM++ after incorporating the RepCaM and TVP modules for long video sequences. For the 1-minute game video, RepCaM achieves an average gain of 0.1 dB compared to the SOTA baseline EMT across all scaling factors. RepCaM++ demonstrates even higher PSNR values, with improvements from 43.72 dB to 43.81 dB ($\times$2) and 34.58 dB to 34.77 dB ($\times$4). For the 5-minute game video, RepCaM consistently provides significant improvements, while RepCaM++ also shows consistent performance boosts, with PSNR values increasing from 43.25 dB to 43.33 dB ($\times$2) and from 34.36 dB to 34.45 dB ($\times$4). These findings highlight the efficacy of the RepCaM and TVPs, where it not only scales up the SR model parameters to enhance perception and manage overfitting with the multi-branch architecture but also maintains scene consistency through various specifically learned TVPs.

\subsection{Application deployment}
One of the many benefits of using RepCaM++ is the ability to utilize smaller SR models with lower capacity and complexity for data overfitting in real-world applications. The patches within each video chunk are often quite similar, particularly in shorter videos, which simplifies the task for smaller models to effectively "memorize" the data. Moreover, unlike CaFM\cite{liu2021overfitting}, the RepCaM++ methods do not require any handcrafted modules, thereby reducing the computational load on end-user devices.

We implemented video chunk processing with associated overfitting models via RepCaM++ on a Huawei P50 Pro, which is powered by a Snapdragon 888 processor, to assess execution performance as shown in \cref{fig:app}. Each video chunk was assigned a unique identifier to streamline its integration into complete frames. We defined real-time performance criteria as having a latency under 500 ms and achieving over 20 FPS on mobile platforms, following the standards set in \cite{zhan2021achieving}. Our method reached 29 FPS while upscaling videos from 270p to 1080p using the EDSR model, markedly outperforming the speed and quality benchmarks of other commonly applied models like EDSR or VDSR, as noted in other baseline studies \cite{liu2021overfitting,li2022efficient}. The deployment results further demonstrate the superiority of our proposed method in both transmission and inference efficiency, as RepCaM++ can consolidate the multi-branch RepCaM to enhance the model perception ability without involving any additional parameter.

\begin{table}[t]
      \begin{center}
        \caption{Ablation Study on the number of parallel branches in the RepCaM module. We adopt EDSR as the backbone and test on $\times$2 and 45s videos.}
            \resizebox{0.475\textwidth}{!}{
        \setlength{\tabcolsep}{1mm}{
       	\begin{tabular}{c|ccccc}
       	\toprule
     Dataset& 1-branch & 2-branch & \cellcolor{green!10}3-branch & 4-branch & 5-branch  \\
        \midrule
        game-45s  & 42.55 & 43.12 & \cellcolor{green!10}43.61 & 43.88 & \textbf{43.96} \\
        vlog-45s  & 48.11 & 48.23 & \cellcolor{green!10}48.48 & 48.61 & \textbf{48.65} \\
        inter-45s & 42.97 & 43.19 & \cellcolor{green!10}43.47 & 43.59 & \textbf{43.64} \\
        sport-45s & 47.97 & 48.25 & \cellcolor{green!10}48.59 & 48.71 & \textbf{48.80} \\
        \bottomrule
		\end{tabular}}
  }
      \label{branch}
      \end{center}
      \vspace{-0.3cm}
\end{table}

\subsection{Comparison with H.264/H.265}
\label{sec:h264}
Our approach can be interpreted as an alternative methodology in video coding. To evaluate its effectiveness, we performed a series of preliminary experiments to benchmark our technique against the established commercial codecs, H.264 and H.265. In these experiments, we manipulated the bitrate of H.264 and H.265 encoded videos to match the size of our LR videos and models while maintaining the original resolution. Subsequently, we assessed the performance of our SR videos against these bitrate-reduced versions of H.264 and H.265 videos. For this purpose, six videos were randomly sampled from the VSD4K dataset. The quantitative outcomes, detailed in \cref{tab:h264}, demonstrate that our method consistently outperforms both H.264 and H.265 across all tested scenarios. For instance, RepCaM++ achieves a PSNR of 43.61 dB, substantially higher than H.264 and H.265 on Game-45s. Similarly, RepCaM++ achieves a PSNR of 48.48  dB on Sport-45s, outperforming H.264 and H.265 by 43.45 dB and 44.24 dB, respectively. These results underscore the superior quality of RepCaM++'s ability to enhance video quality compared to traditional codec standards.


\subsection{Qualitative results} 
\label{sec:visual}
We present qualitative results for our proposed RepCaM++ method and conduct a comparative analysis of video super-resolution outcomes across the game, sport, vlog, and inter datasets, as shown in \cref{fig:vis}. RepCaM++ consistently outperforms the CaFM and EMT baselines, demonstrating superior detail preservation and clarity. Specifically, in the Game and Vlog datasets, RepCaM++ distinctly renders numbers and text that are indiscernible with the baseline methods. In the Sport dataset, our method captures intricate skateboard patterns, while baselines produce blurring. In the Inter dataset, RepCaM++ delineates fine contours of small jewelry, which the baselines fail to maintain. These results highlight the enhanced capability of RepCaM++ in preserving fine-grained details and textual clarity, especially when integrating TVPs to specifically perceive the foreground objects for more detailed information.

\section{Conclusion and future research}
\label{conclusion}
We explore a new perspective of content-aware modulation methods for neural video delivery termed RepCaM++. The proposed RepCaM and TVP methods compensate each other to fit integrated video and accelerate the end-end training with less than 0.01\% additional overhead. During training, our method adopts extra parallel-cascade parameters to enhance the generalization ability of the super-resolution model and overfit multiple video chunks while removing the additional parameters through re-parameterization during inference.

For future research, we aim to reduce transmission overheads further in two directions. First, we plan to compress delivered content-aware models by designing quantization methods tailored for such models, enabling 4-bit or smaller models to retain content awareness. Pruning is another promising approach to reduce training costs without sacrificing performance. Second, we intend to discard redundant frames during video chunk transmission and apply frame interpolation algorithms on the client side for restoration. However, effective filtering of redundant frames requires further investigation.


\bibliographystyle{IEEEtran}
\bibliography{mybibfile}

\begin{IEEEbiography}[{\includegraphics[width=1in,height=1.25in,clip,keepaspectratio]{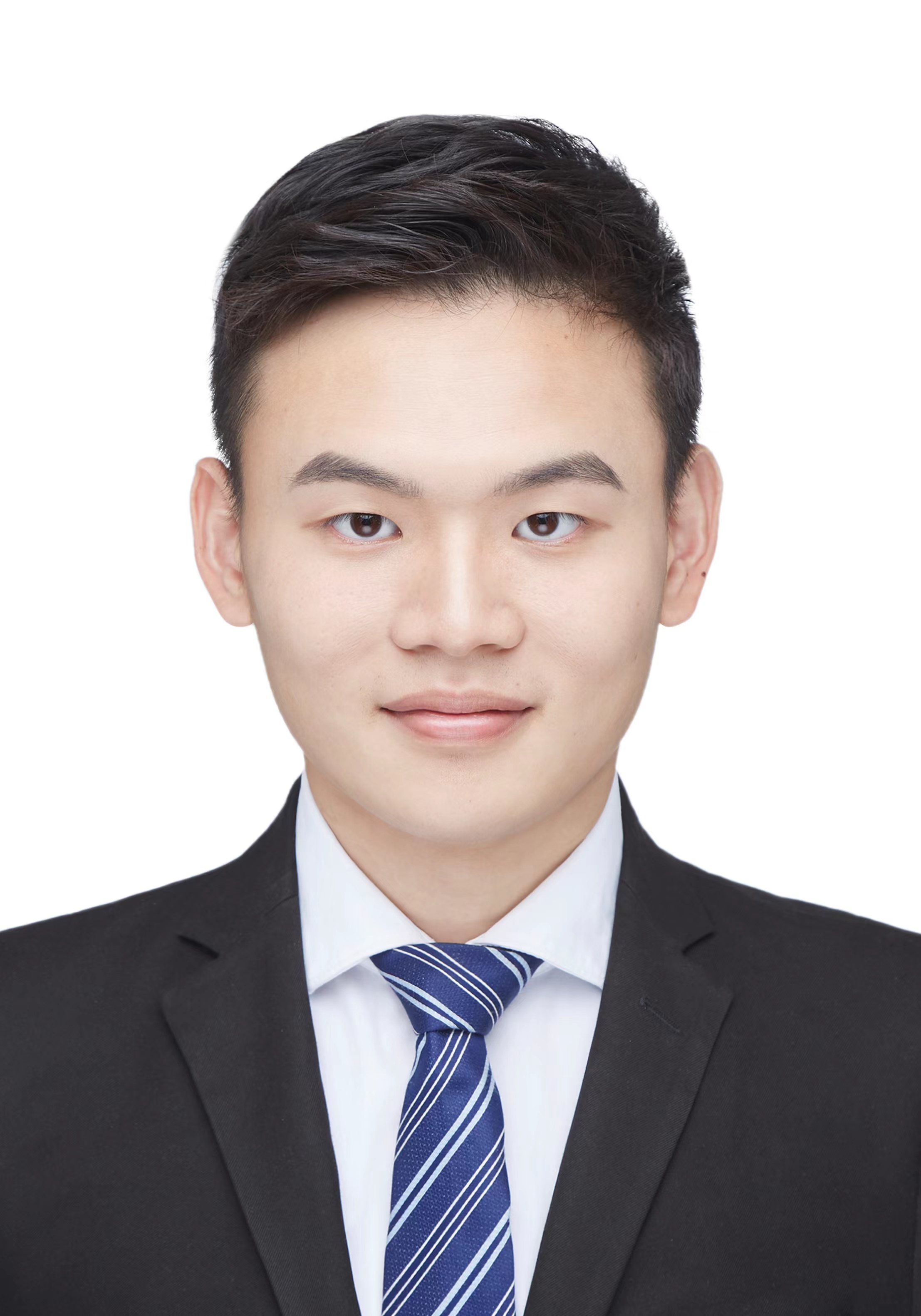}}]{Rongyu Zhang}
is a dual Ph.D. student at Nanjing University and The Hong Kong Polytechnic University. He received M.Phil. degree at The Chinese University of Hong Kong, Shenzhen in 2023. He also received both B.E. degree and B.M. degree from the joint program of Beijing University of Posts and Telecommunications and Queen Mary University of London in 2021. His research interests include Federated Learning and Efficient Learning for LLM and VLM.

\end{IEEEbiography}

\vskip -2\baselineskip plus -1fil

\begin{IEEEbiography}[{\includegraphics[width=1in,height=1.25in,clip,keepaspectratio]{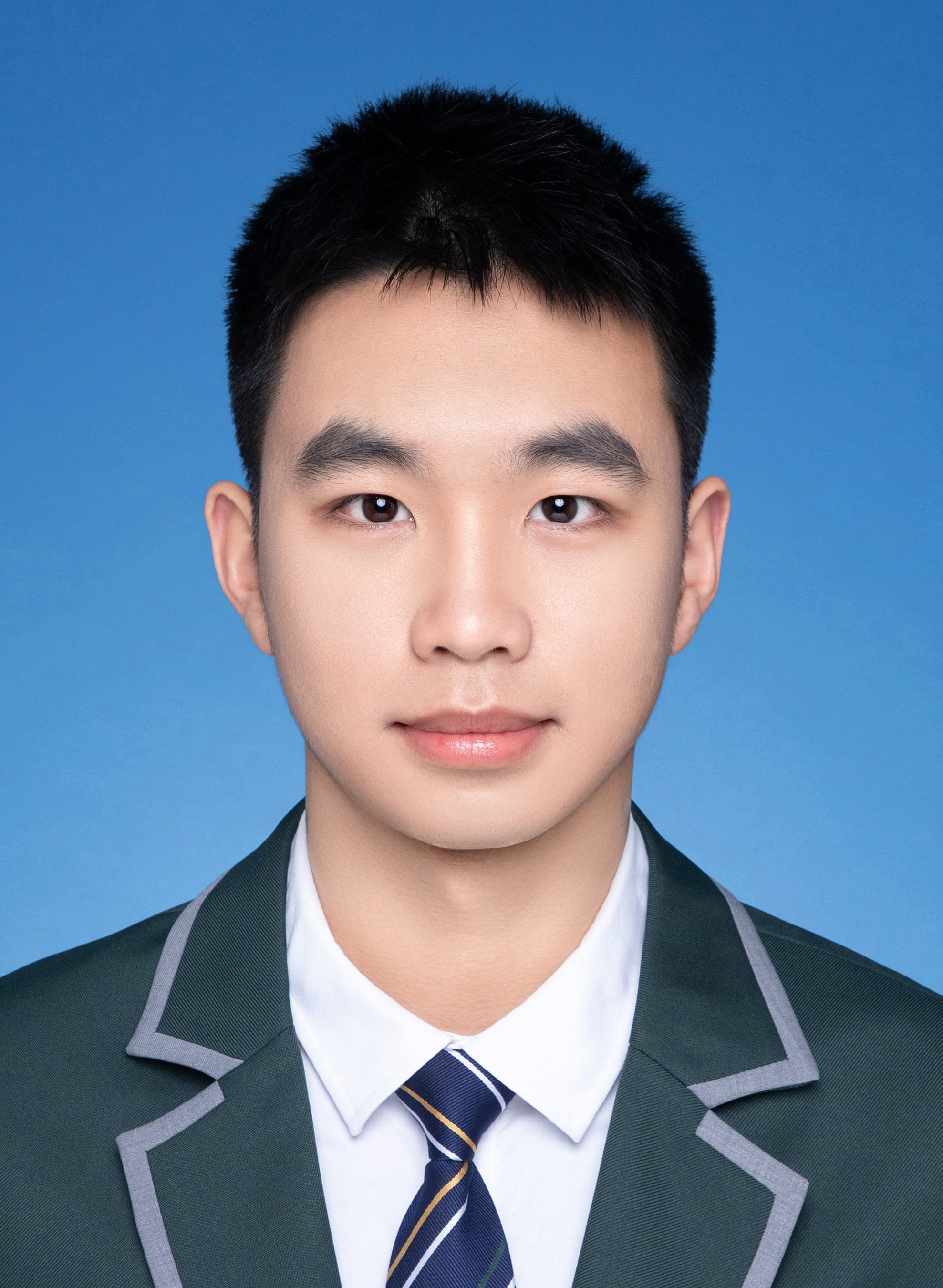}}]{Xize Duan}
is an undergraduate student at The Chinese University of Hong Kong, Shenzhen, studying at the School of Science and Engineering (SSE). Currently, he serves as a research assistant in INML under the guidance of Professor Fangxin Wang. He is currently interested in computer vision, multimedia and network transmission.

\end{IEEEbiography}

\vskip -2\baselineskip plus -1fil

\begin{IEEEbiography}[{\includegraphics[width=1in,height=1.15in,clip,keepaspectratio]{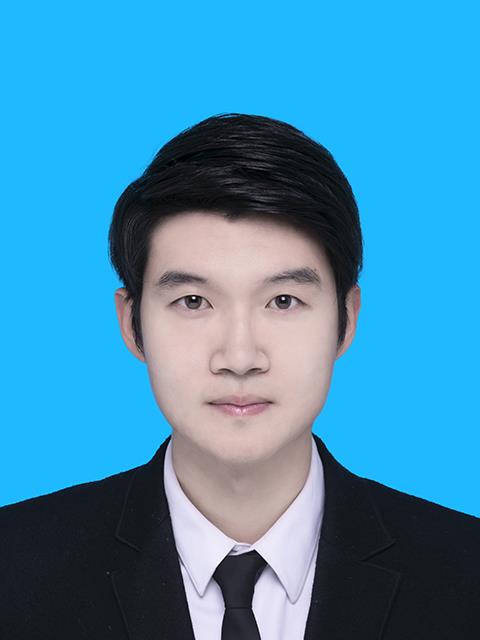}}]{Jiaming Liu}
 received the B.S. and M.S. degrees in Information and Communication Engineering from Beijing University of Posts and Telecommunications, China, in 2019 and 2022, receptively. Currently, he is pursuing his Ph.D. degree in Computer Science and Technology at Peking University. His research interests include out-of-distribution generalization, autonomous driving, and multimodal scene understanding. \end{IEEEbiography}

\vskip -2\baselineskip plus -1fil

\begin{IEEEbiography}[{\includegraphics[width=1in,height=1.1in,clip,keepaspectratio]{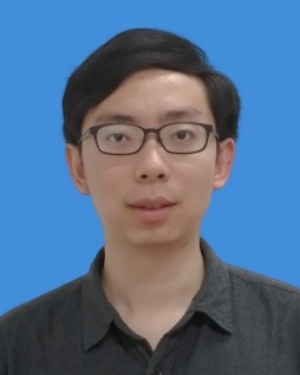}}]{Li Du}
(M’16) received his B.S degree from Southeast University, China and his Ph.D. degree in Electrical Engineering from University of California, Los Angeles.  Currently, he is an associate professor in the department of Electrical Science and Engineering at Nanjing University. His research includes analog sensing circuit design, in-memory computing design and high-performance AI processor for edge sensing.\end{IEEEbiography}

\vskip -2\baselineskip plus -1fil

\begin{IEEEbiography}[{\includegraphics[width=1in,height=1.1in,clip,keepaspectratio]{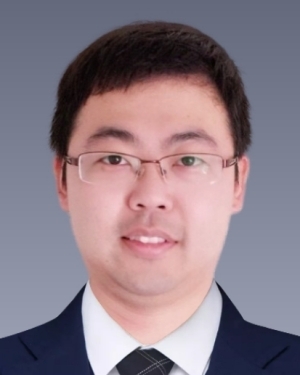}}]{Yuan Du}
(S’14-M’17-SM’21) received his B.S. degree from Southeast University (SEU), Nanjing, China, in 2009, his M.S. and his Ph.D. degree both from Electrical Engineering Department, University of California, Los Angeles (UCLA), in 2012 and 2016, respectively. Since 2019, he has been with Nanjing University, Nanjing, China, as an Associate Professor. His current research interests include designs of machine-learning hardware accelerators, and RFICs.
\end{IEEEbiography}

\vskip -2\baselineskip plus -1fil

\begin{IEEEbiography}[{\includegraphics[width=1in,height=1.15in,clip,keepaspectratio]{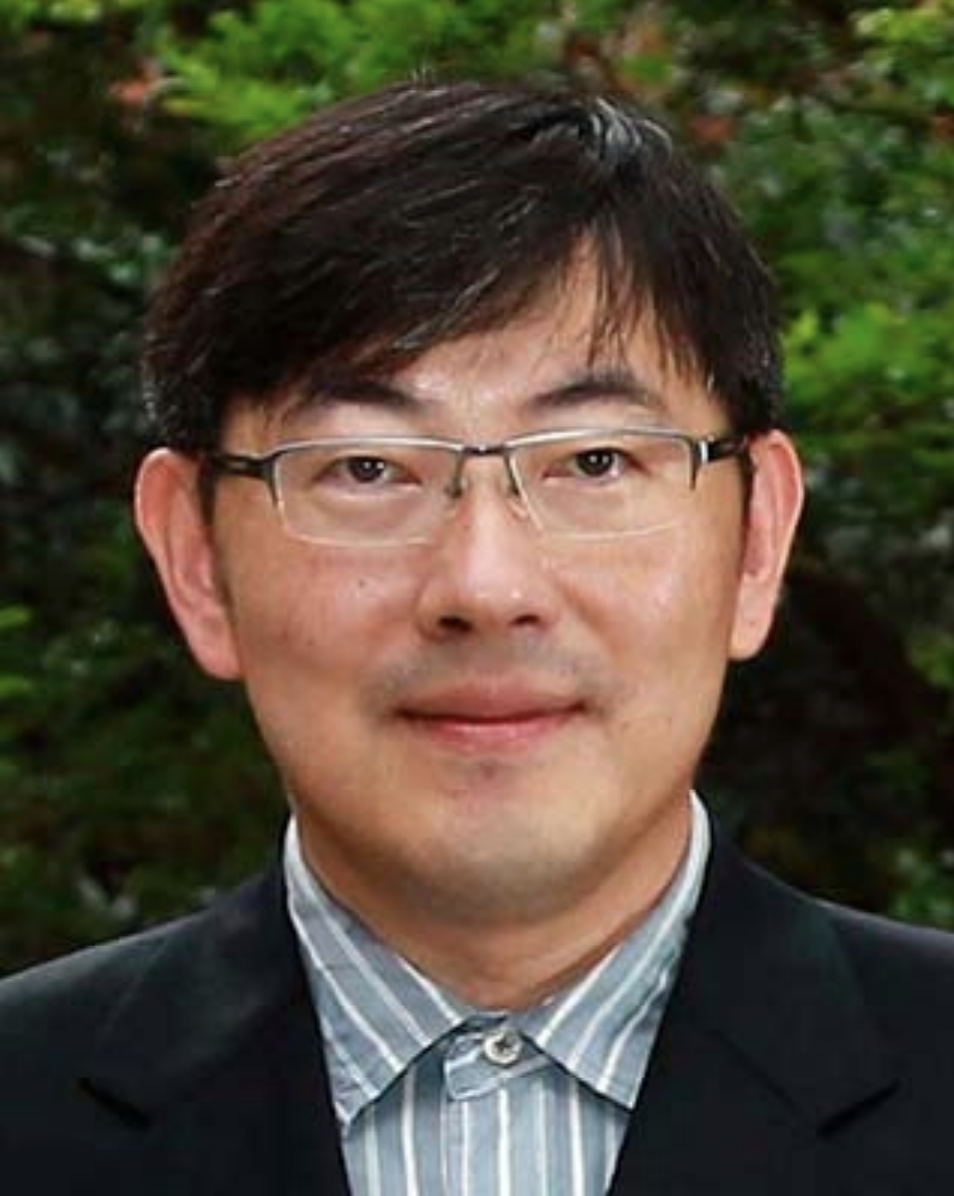}}]{Dan Wang}
is a Professor of Department of Computing, The Hong Kong Polytechnic University. He received the Ph.D. degree from Simon Fraser University. His research falls in general computer networking and systems, where he published in ACM SIGCOMM, ACM SIGMETRICS, and the IEEE INFOCOM, and many others. He is the Steering Committee Chair of IEEE/ACM IWQoS. His research interests include network architecture and QoS, smart building, and Industry 4.0.
\end{IEEEbiography}

\vskip -2\baselineskip plus -1fil

\begin{IEEEbiography}[{\includegraphics[width=1in,height=1.1in,clip,keepaspectratio]{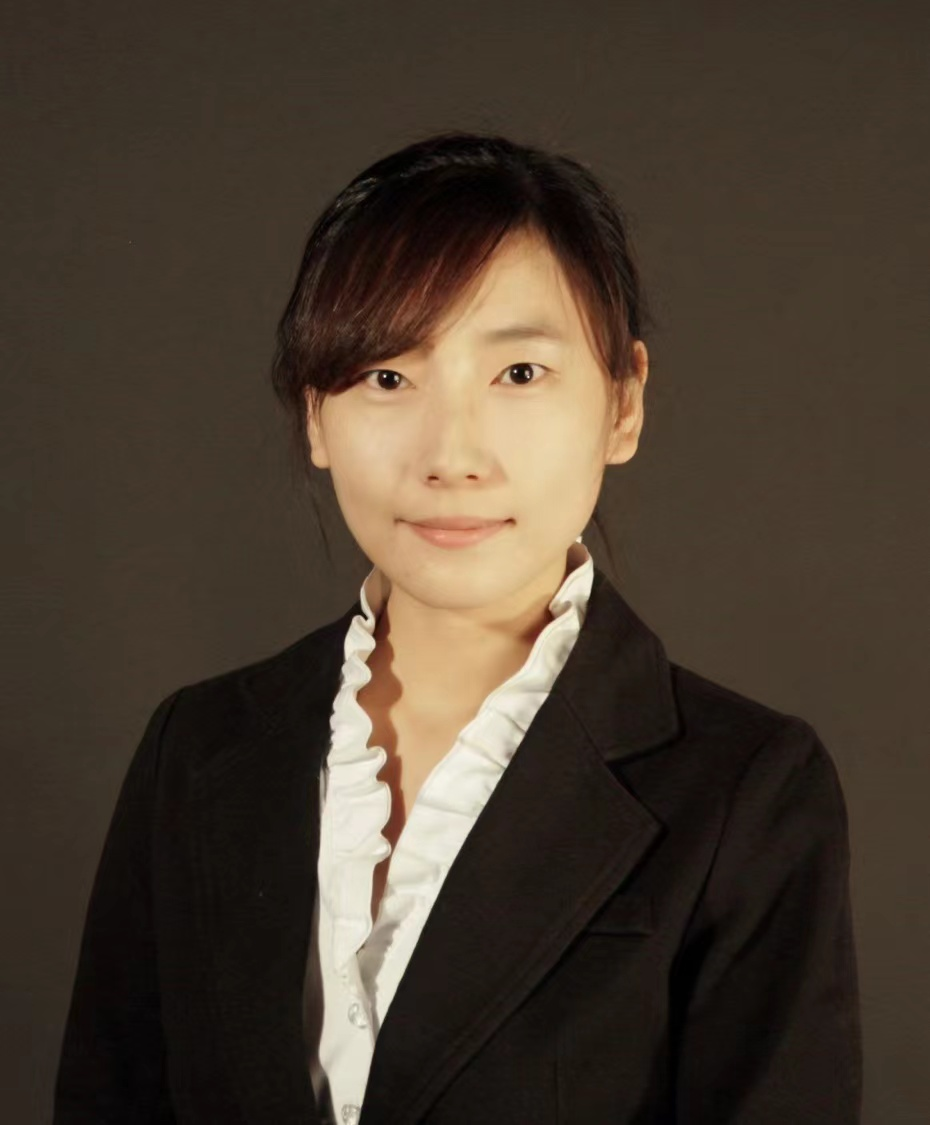}}]{Shanghang Zhang}
received her M.S. degree from Peking University, and her Ph.D. degree from Carnegie Mellon University in 2018. After that, she has been the postdoc research fellow at Berkeley AI Research Lab, UC Berkeley. Currently, she's an assistant professor at the School of Computer Science, Peking University. Her research focuses on machine learning generalization in the open world, including theory, algorithm, and system development, with applications to important IoT problems.
\end{IEEEbiography}

\vskip -2\baselineskip plus -1fil

\begin{IEEEbiography}[{\includegraphics[width=1in,height=1.25in,clip,keepaspectratio]{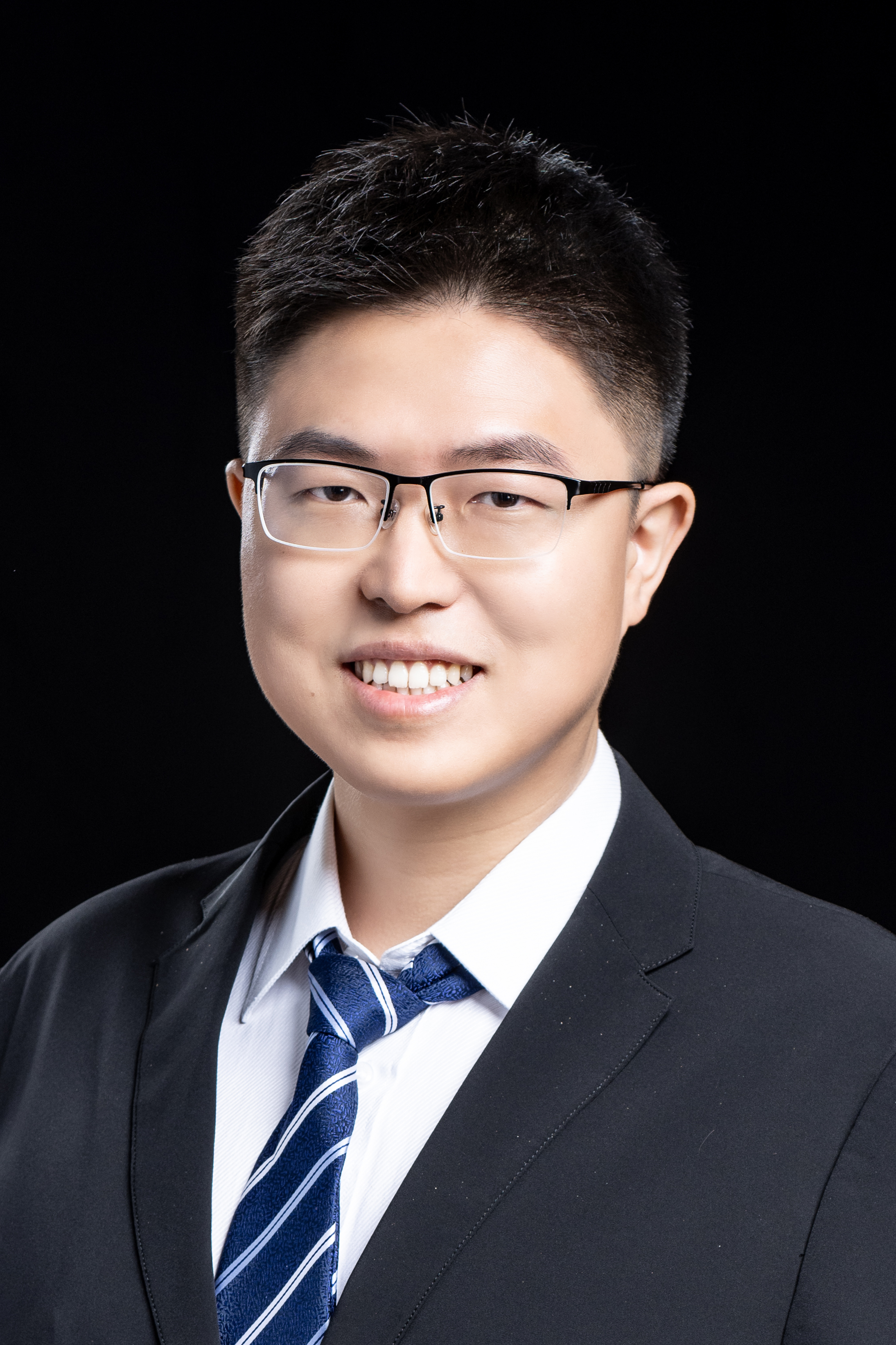}}]{Fangxin Wang}(S'15-M'20) is an assistant professor at The Chinese University of Hong Kong, Shenzhen (CUHKSZ). He received the Ph.D., M.Eng., and B.Eng. degree all in Computer Science and Technology from Simon Fraser University, Tsinghua University, and Beijing University of Posts and Telecommunications, respectively. Before joining CUHKSZ, he was a postdoctoral fellow at the University of British Columbia. Dr. Wang's research interests include Multimedia Systems and Applications, Cloud and Edge Computing, Deep Learning and Big Data Analytics, Distributed Networking and System. He leads the intelligent networking and multimedia lab at CUHKSZ. He has published more than 30 papers at top journal and conference papers, including INFOCOM, Multimedia, ToN, TMC, IOTJ, etc. He served as the publication chair of IEEE/ACM IWQoS, TPC member of IEEE ICC, and reviewer of many top conference and journals, including INFOCOM, ToN, TMC, IOTJ.
\end{IEEEbiography}

\newpage

\vfill

\end{document}